\documentclass[twocolumn,showpacs,prb,superscriptaddress,amsmath,floatfix]{revtex4}
 
\usepackage{graphicx}

\newcommand{\figurewidth}{\columnwidth}

\begin{document}

\title{Critical behavior of the three-dimensional bond-diluted Ising
  spin glass:\\ Finite-size scaling functions and universality}
\author{Thomas J\"{o}rg} \affiliation{Dipartimento di Fisica,
  Universit\`a di Roma ``La Sapienza'', SMC and INFM, P.le.~Aldo Moro
  2, 00185 Roma, Italy.}

\date{\today}

\begin{abstract}
  We study the three-dimensional (3D) bond-diluted Edwards-Anderson
  (EA) model with binary interactions at a bond occupation of $45\%$
  by Monte Carlo (MC) simulations. Using an efficient cluster MC
  algorithm we are able to determine the universal finite-size scaling
  (FSS) functions and the critical exponents with high statistical
  accuracy.  We observe small corrections to scaling for the measured
  observables.  The critical quantities and the FSS functions indicate
  clearly that the bond-diluted model for dilutions above the critical
  dilution $p^*$, at which a spin glass (SG) phase appears, lies in
  the same universality class as the 3D undiluted EA model with binary
  interactions. A comparison with the FSS functions of the 3D
  site-diluted EA model with Gaussian interactions at a site
  occupation of $62.5\%$ gives very strong evidence for the
  universality of the SG transition in the 3D EA model.

\end{abstract}


\maketitle

\section{Introduction}

Almost all the Monte Carlo (MC) simulations of the three-dimensional
(3D) Edwards-Anderson (EA) model\cite{edwards:75} have been performed
using either an undiluted Gaussian or binary coupling
distribution.\cite{diep_cha:05}  There are, however, many reasons to
study the 3D bond-diluted EA model with binary interactions. The
simplest reason being that it allows one, although in a very crude way, to
approximate the presence of weak bonds in the model with Gaussian
couplings without loosing the technical benefits of the binary
couplings. The introduction of vacant bonds reduces the frustration in
the model, making it easier to find ground states, which should
facilitate low-temperature MC simulations.  Moreover, the bond-diluted
model in a certain range of the dilution allows for an efficient use
of replica cluster MC algorithms.\cite{swendsen:86, redner:98,
  houdayer:01, wang:05, jorg:05} It has been shown that the 
high-temperature series expansion typically converges better for some
intermediate dilution than for the undiluted case.\cite{shapira:94}
Similarly, it was found that the scaling of the ground state defect
energies is much improved at intermediate dilutions in comparison to
the standard undiluted case, that showed to scale rather
poorly.\cite{boettcher:04}  This gives some evidence that in general
the scaling violations of the diluted model, for some intermediate
range of the dilution, might be smaller than in the undiluted case.
This fact is very important for spin glass (SG) simulations since the
range of system sizes that are accessible to MC calculations is very
limited due to the presence of large free energy barriers in the SG
phase, causing slow thermalization. The problems of the very
restricted simulation range are worsened by the fact that the leading
correction-to-scaling exponent $\omega$ has been measured to be
probably smaller than one,\cite{ballesteros:00, palassini:99b} such
that the leading corrections fall off only slowly. All this combined
with the need of excellent statistics makes a reliable estimate of
critical exponents very difficult.  In addition to these rather
technical aspects the bond-diluted $\pm J$ EA model may be thought of
simply as an EA model with a particular distribution of the couplings
and therefore it also serves to provide insights on the debated issue
of the universality of the 3D EA model.\cite{bernardi:96, bernardi:97,
  mari:02, daboul:04, pleimling:05,katzgraber:06,toldin:06}

In this paper we first consider in detail the bond-diluted 3D EA
model at a bond occupation of $45\%$ which is close to the value where
optimal scaling for the ground state defect energies was
found.\cite{boettcher:04} We first discuss the replica cluster MC algorithm
we use and its performance, then we focus on the SG transition, the
corresponding critical exponents and the universal finite-size scaling
(FSS) functions of the correlation length $\xi$, the SG susceptibility
$\chi_{\rm SG}$ and the Binder ratio $g$. We determine these FSS functions
with very high statistical accuracy (up to $10^5$ samples per lattice
size) and use them to determine critical exponents with two different
methods, namely, the quotient method
\cite{ballesteros:97,ballesteros:00} and an extrapolation to infinite
volume method.\cite{caracciolo:95} We find excellent agreement
between the critical exponents extracted by these two different
methods, as well as with the values given by the high-precision study
of the undiluted binary model of Ballesteros {\it et
al}.\cite{ballesteros:00}  In addition, we find that the FSS functions
agree very well with the ones obtained by Palassini and Caracciolo for
the undiluted binary model.\cite{palassini:99b} This together gives
strong evidence that the SG transition of the 3D bond-diluted EA model
with binary couplings falls into the same universality class as the
one of the corresponding undiluted model, as indicated by high-temperature
expansion\cite{shapira:94} and as one might expect on general
grounds.\cite{ harris:74} We then compare the FSS functions to the
ones of the 3D site-diluted EA model at a site occupation of $62.5\%$
with a Gaussian distribution of the couplings.\cite{jorg:05} In
contrast to the bond-diluted $\pm J$ model the site-diluted Gaussian
model shows very noticeable finite-size effects and therefore the
agreement between the FSS functions of the two models emerges only at
large enough sizes. Despite the fact that the two models have
finite-size effects of clearly different magnitude the agreement
between the FSS functions gets excellent asymptotically. This leads us
to the conclusion that the critical behavior of the 3D EA model is
most probably universal and that violations of universality as they
are found in dynamical MC simulations\cite{bernardi:96, bernardi:97,
  mari:02, pleimling:05} are probably due to difficulties to control
the corrections to scaling in such
simulations.\cite{parisi:99x,jimenez:06} In
Ref.~\onlinecite{katzgraber:06} the conclusion that the critical
behavior of the 3D EA model is universal is reached from a high-precision 
MC study in which the static properties of the model with Gaussian 
and $\pm J$ couplings are compared.

We find indications that scaling violations at the given bond
occupation are in fact reduced in comparison to the standard undiluted
model, making the study of the bond-diluted $\pm J$ EA very interesting
for various reasons.

Finally, note that all the figures in this paper where the coupling
distribution is not explicitly mentioned in the caption refer to the
data obtained from the study of the bond-diluted $\pm J$ EA model.

\section{The Models} 

\subsection{Bond-diluted $\pm J$ EA model}
We mainly consider the 3D bond-diluted EA model given by the
Hamiltonian\cite{edwards:75}
\begin{equation}
  \label{eq:diluted_EA}
  {\cal H} = - \sum_{\langle xy \rangle} J_{xy} \epsilon_{xy} \sigma_x \sigma_y \,,
\end{equation}
where the sum is over all the nearest neighbor pairs of a simple cubic
3D lattice of length $L$. The $\sigma_x = \pm 1$ are Ising spins and
the $\epsilon_{xy}$ are the bond occupation variables,
i.e.,~$\epsilon_{xy} = 0$ for an empty and $\epsilon_{xy} = 1$ for an
occupied bond. For the couplings $J_{xy}$ we use a binary
distribution, i.e.,~$J_{xy} = \pm 1$. The fraction of occupied bonds
$p$ in this study is $p = 0.45$. We use periodic boundary conditions
to study the Hamiltonian in Eq.~(\ref{eq:diluted_EA}).  Note that a SG
transition occurs only at and above a certain critical dilution $p^*$
that is found to be slightly larger than the bond percolation
threshold $p_c = 0.2488$ for discrete distributions of the couplings.
\cite{brayfeng:87,boettcher:04} This is in contrast to the case of
continuous coupling distributions, where the SG phase appears exactly
at $p_c$.  A detailed discussion on the value of $p^*$ will be given
in Ref.~\onlinecite{jimenez:06}.

\subsection{Site-diluted Gaussian EA model}
In order to compare the FSS functions for two different coupling
distributions we also consider the 3D site-diluted EA model with a
Gaussian coupling distribution given by the
Hamiltonian\cite{edwards:75}
\begin{equation}
  \label{eq:diluted_EA_gauss}
  {\cal H} = - \sum_{\langle xy \rangle} J_{xy} \epsilon_{x} \epsilon_{y} \sigma_x \sigma_y \,,
\end{equation}
where the sum is again over all the nearest neighbor pairs of a simple
cubic 3D lattice of length $L$. The $\sigma_x = \pm 1$ are Ising spins
and the $\epsilon_{x}$ are the site occupation variables, i.e.,
$\epsilon_{x} = 0$ for an empty and $\epsilon_{x} = 1$ for an occupied
site. For the couplings $J_{xy}$ we use a Gaussian distribution with
mean zero and variance unity. The fraction of occupied sites $r$ in
this study is $r = 0.625$. For this model the SG phase is supposed to
be present above the site percolation threshold $r_c = 0.3116$.  We
again use periodic boundary conditions to study the Hamiltonian in
Eq.~(\ref{eq:diluted_EA_gauss}).

\section{The FSS method}

Taking a suitably defined finite-volume correlation length $\xi(T,L)$
and a long-range observable ${\cal{O}}(T,L)$ as, e.g., $\chi_{\rm SG}(T,L)$
FSS theory predicts\cite{fisher:72,privman:90}
\begin{equation}
  \label{eq:FSS_1}
  \frac{{\cal O}(T,L)}{{\cal O}(T,\infty)}  =
  f_{\cal O} \bigl[ \xi(T,\infty)/L \bigr] + O\Bigl(\xi^{-\omega},L^{-\omega}\Bigr)
\end{equation}
and
\begin{equation}
  \label{eq:FSS_2}
  \frac{{\cal O}(T,s L)}{{\cal O}(T,L)}   =
  F_{{\cal O}} \bigl[ \xi(T,L)/L; s \bigr] + O\bigl(\xi^{-\omega},L^{-\omega}\Bigr)
\end{equation}
where $f_{{\cal O}}$ and $F_{{\cal O}}$ are universal FSS functions
and $s > 1$ is a scale factor.  Eq.~(\ref{eq:FSS_2}) is an excellent
starting point for numerical investigations of the FSS behavior, as it
involves only finite-volume quantities taken from a pair of systems
with sizes $L$ and $s L$ at a given temperature $T$. The knowledge of
the universal scaling functions $F_{{\cal O}}$ and $F_{{\xi}}$ allows
us to extract critical exponents either by infinite volume
extrapolation \cite{caracciolo:95, palassini:99b} or by the quotient
method.\cite{ballesteros:97,ballesteros:00}  For the quotient method
one defines an effective critical temperature $T^*_c$ at which the
correlation length measured in units of the lattice size $L$ are equal
for the pair of systems, i.e.,
\begin{equation}
  \label{eq:equal_correlation_length}
  \xi(T^*_c,L)/L = \xi(T^*_c,sL)/(sL) \,.  
\end{equation}
For an observable ${\cal{O}}$ that in the thermodynamic limit diverges
as $t^{-x_{\cal O}}$ ($t=T/T_c-1$ being the reduced temperature) the
critical exponent $x_{\cal O}$ can be extracted from the quotient
\begin{equation}
  \label{eq:quotient_1}
  s^{x_{\cal O}/\nu} = \frac{{\cal O}(T^*_c,s L)}{{\cal O}(T^*_c,L)} + O(L^{-\omega}) 
\end{equation}
and hence from $F_{{\cal O}}$. The critical exponent $\nu$ describing
the divergence of $\xi$ can be, e.g., measured from the FSS function of
the temperature derivative of $\xi$, i.e., from $F_{{\cal
    \partial_{\rm T} \xi}}$, which after a short calculation leads to
\begin{equation}
  \label{eq:quotient_2}
  s^{1/\nu} = 1 + \frac{x^*}{s} \partial_x F_{{\cal \xi}} (x;s) \big|_{x=x^*} + O(L^{-\omega}) \,  
\end{equation}
with $x = \xi(T,L)/L$ and $x^* = \xi(T_c^*,L)/L$.  The infinite volume
extrapolation technique works with data strictly above $T_c$ and uses
the FSS functions $F_{{\cal \xi}}$ and $F_{{\cal O}}$ to obtain the
thermodynamic limit of ${\cal O}$ using an iterative procedure in
which the pair $\xi$ and ${\cal O}$ is scaled up from $L \to s L \to
s^2 L \to \ldots \to \infty$ as described in detail in
Ref.~\onlinecite{caracciolo:95}.  Assuming an ordinary second-order
phase transition $x_{\cal O}$ can be extracted from a power-law fit to
the extrapolated data of ${\cal O}(T,\infty)$.

\section{Algorithm and computational details}
In this MC study we make use of the fact that the Hamiltonians in
Eqs.~(\ref{eq:diluted_EA}) and (\ref{eq:diluted_EA_gauss}) allow for
an efficient use of replica cluster algorithms that were originally
used for the study of the two-dimensional EA model.\cite{swendsen:86,
  houdayer:01, wang:05}  The details of the algorithm as they are
given below apply to the case of the bond-diluted $\pm J$ model,
whereas a discussion of the algorithm in the case of the site-diluted
Gaussian case can be found in Ref.~\onlinecite{jorg:05}.  The basic
idea of all these algorithms is that two independent replicas of the
system are simulated simultaneously. Hence each lattice site can be
associated with two spins, one from each replica, and can be in one of
the four spin states $(++)$, $(+-)$, $(-+)$, and $(--)$. The $(+-)$
and $(-+)$ sites are called A sites and the $(++)$ and $(--)$ sites B
sites. On these two different groups of sites it is now possible to
define clusters in various ways. In the following we define the
cluster moves used for this simulation.  First we consider the
clusters on the A sites. Following Redner {\it et al}.~\cite{redner:98} we
grow the clusters by adding links between two A sites with probability
$p_{add}$ given by
\begin{equation}
  \label{eq:link_addition}
  p_{\rm add} =
  \begin{cases} 
    1 -\exp(4 \beta E_{xy})\quad {\rm if} \quad E_{xy} < 0 \\
    0\quad \rm{else}, \\
  \end{cases}
\end{equation}
where $E_{xy} = - J_{xy} \epsilon_{xy} \sigma_x \sigma_y$ is the
energy contribution of the given link and $\beta$ is the inverse
temperature. We repeat the same procedure for the links between the B
sites. Finally, we flip each of the clusters that are defined through
this procedure with probability 1/2, i.e., we make a Swendsen-Wang
update.\cite{swendsen:87} This choice is influenced by the findings
that even in a diluted ferromagnet an update of all clusters can be
efficient.\cite{ballesteros:98}  In the presence of an external field
the algorithm would have to be modified slightly as in this case only
the clusters on the A sites flip freely.\cite{redner:98,houdayer:01}

In order to improve the performance of this cluster algorithm we also
update each replica with parallel tempering (PT)
(Ref.~\onlinecite{hukushima:96}) as proposed by Houdayer.\cite{houdayer:01} 
This means that in addition to
the two replicas at temperature $T$ we introduce replicas on a set of
$N_T$ temperatures ($T_i$ with $i=1, \ldots, N_T$) and allow two
replicas at $T_i$ and $T_{i+1}$ to exchange their temperatures with
the appropriate probability.\cite{hukushima:96}  The resulting
algorithm is only a slight variation of the original replica cluster
algorithm by Swendsen and Wang.\cite{swendsen:86} In contrast to
Houdayer's algorithm\cite{houdayer:01} it does not need PT to ensure
ergodicity, because the different definition of the clusters already
provides for it.  The dilution in the coupling distribution is very
important in order to avoid the site percolation problem mentioned by
Houdayer.\cite{houdayer:01}  Using, e.g., a nondiluted $\pm J$
coupling distribution the cluster move essentially just swaps the two
replicas. That dilution may help greatly to speed up cluster algorithms
in SG simulations has already been noted in the case of the Viana-Bray
model\cite{viana:85,persky:96}.  Finally, we note that it would also
be possible to define clusters along the lines of
Ref.~\onlinecite{niedermayer:88}, where the size of the resulting
clusters can be controlled in a more general manner.

\begin{figure}[tbf]
  \includegraphics[width=\figurewidth]{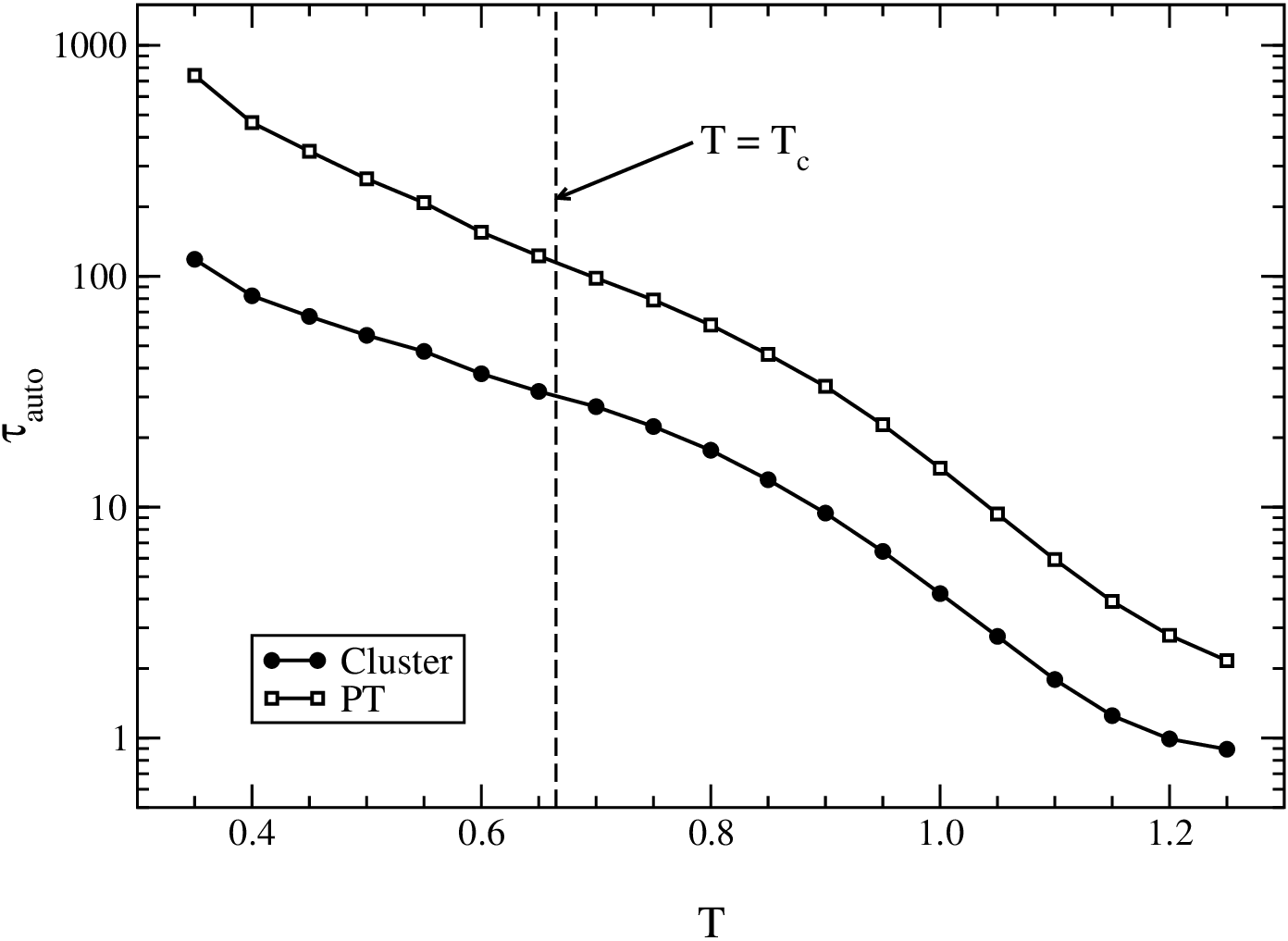}
  \caption{\label{fig:auto_time} Qualitative comparison of the integrated 
    autocorrelation time $\tau_{\rm auto}$ as a function of
    temperature between parallel tempering (PT) and the replica
    cluster algorithm (Cluster) averaged on the same 20 configurations
    at $L=10$.}
\end{figure}

An important technical advantage of this algorithm with the given
discrete distribution of the couplings is that it can be completely
multispin coded, apart from the task where the clusters are identified
with the Hoshen-Kopelman algorithm.\cite{hoshen:76}  In
Fig.~\ref{fig:auto_time} we compare the integrated autocorrelation
times of the square of the order parameter $q^2$ (as defined below) of
this cluster algorithm with the corresponding ones of PT with the same
number of replicas simulated in total, i.e., with the same number of
temperature slices, the same set of temperatures, and two replicas at
each temperature. For this qualitative comparison we show an average
over the same 20 configurations on $L=10$ lattices. One sees that in
the SG phase the autocorrelation time of the cluster algorithm for the
particular choice of temperatures and size is one order of magnitude
smaller than the one of PT. The computational overhead of the cluster
move can be kept rather small, such that the overall performance
remains clearly favorable for the cluster algorithm.

We measure the overlap $q_x = \sigma_x \tau_x$ and $q = L^{-3}\sum_x
q_x $ from two replicas $\sigma$ and $\tau$ with the same couplings
$J_{xy}\epsilon_{xy}$.  We define $\xi(T,L)$ by the second-moment
correlation
length\cite{cooper:82,kim:94,caracciolo:95,palassini:99b,ballesteros:00}
\begin{equation}
  \label{eq:correlation_length}
  \xi(T,L) = \frac{1}{2 \sin(|{\mathbf k}_{\rm min}|/2)} \left[
    \frac{\chi_{\rm SG}(\mathbf 0)}{\chi_{\rm SG}({\mathbf k}_{\rm min})} - 1 \right]^{\frac{1}{2}} \,,
\end{equation}
where the wave-vector-dependent SG susceptibility is defined by
\begin{equation}
  \label{eq:susceptibility}
  \chi_{\rm SG}({\mathbf k}) = \frac{1}{L^3} \sum_{\mathbf x}\sum_{\mathbf r} e^{i {\mathbf k \mathbf r}} 
  [\langle q_{\mathbf x} q_{\mathbf x+ \mathbf r} \rangle_{\rm T} ]_{\rm av}
\end{equation}
and where ${\mathbf k}_{\rm min} = (0,0,2 \pi/L)$ is the smallest non-zero
wavevector allowed by periodic boundary conditions. The SG
susceptibility is defined through $\chi_{\rm SG}(T,L) = L^3 \langle q^2
\rangle = \chi_{\rm SG}(\mathbf 0)$. We denote the thermal average at
temperature $T$ by $\langle \ldots \rangle_{\rm T}$ and the average over
the disorder realizations by $[ \ldots ]_{\rm av}$. We also measure
the Binder ratio defined by\cite{binder:81,binder:81b}
\begin{equation}
  \label{eq:binder}
  g(T,L) = \frac{1}{2} \left( 3 - \frac{[\langle q^4\rangle_{\rm T}]_{\rm av}}{[\langle q^2\rangle_{\rm T}]_{\rm av}^2}\right)\, .
\end{equation}

The parameters used for the simulation of the bond-diluted $\pm J$ EA
model are given in Table \ref{tab:simulation_1}, whereas the ones used
for the simulation of the site-diluted Gaussian EA model are given
Table \ref{tab:simulation_2}.  The number of replicas used at each
temperature is two. The number of MC sweeps for the measurements is
equal to the number of sweeps used for thermalization. Thermalization
has been assured by requiring that the observables agree within errors
in the last three bins of a logarithmic binning. The statistical
errors of the MC observables are obtained from a jackknife
estimate.\cite{efron:82}

\begin{table}
  \begin{ruledtabular}
    \begin{center}
      \begin{tabular}{rrrrc}
        $L$ & $N_{s}$ & $N_{\rm therm}$ & $N_T$ & $T$ \\
        \hline
        4  &  50240 &  20000 & 19 & 0.50 -- 1.40 \\
        5  &  50200 &  20000 & 19 & 0.50 -- 1.40 \\
        6  &  50200 &  20000 & 19 & 0.50 -- 1.40 \\
        7  &  50240 &  20000 & 19 & 0.50 -- 1.40 \\
        8  & 102772 &  20000 & 19 & 0.50 -- 1.40 \\
        9  &  51200 &  40000 & 19 & 0.50 -- 1.40 \\
        10 &  51200 &  30000 & 19 & 0.50 -- 1.40 \\
        12 &  40320 &  64000 & 19 & 0.50 -- 1.40 \\
        15 &   5024 & 100000 & 49 & 0.55 -- 1.40 \\
        16 &  10496 & 200000 & 33 & 0.60 -- 1.40 \\
      \end{tabular}
    \end{center}
  \end{ruledtabular}
  \caption{\label{tab:simulation_1} Simulation parameters for the 
    bond-diluted $\pm J$ EA model. Number of samples $N_{s}$, minimal 
    number of Monte Carlo sweeps for thermalization $N_{\rm therm}$, 
    the temperature range $T$, and the number of temperatures $N_T$ 
    used for the PT as a function of the size $L$.}
\end{table}  

\begin{table}
  \begin{ruledtabular}
    \begin{center}
      \begin{tabular}{rrrrc}
        $L$ & $N_{s}$ & $N_{\rm therm}$ & $N_T$ & $T$ \\
        \hline
        4  & 8000 &  8000 & 21 & 0.40 -- 1.40 \\
        6  & 8000 & 10000 & 21 & 0.40 -- 1.40 \\
        8  & 8000 & 12000 & 21 & 0.40 -- 1.40 \\
        10 & 4705 & 15000 & 21 & 0.40 -- 1.40 \\
        12 & 4000 & 25000 & 21 & 0.40 -- 1.40 \\
        16 & 3639 & 40000 & 21 & 0.40 -- 1.40 \\
        20 &  333 & 90000 & 21 & 0.40 -- 1.40 \\
      \end{tabular}
    \end{center}
  \end{ruledtabular}
  \caption{\label{tab:simulation_2} Simulation parameters for the 
    site-diluted Gaussian EA model. The meaning of the parameters 
    is the same as in Table \ref{tab:simulation_1}.}
\end{table}

\section{FSS analysis and results}

In the following we make a detailed FSS analysis of the bond-diluted
$\pm J$ EA model and present results on its critical behavior. After
that we discuss a few results on the site-diluted Gaussian EA model
which are mainly to motivate that in this model the finite-size
effects are strong. Its FSS behavior is discussed in more detail in
Sec.~\ref{sec:universality}, where we make a comparison between the
FSS functions of the two models. An independent determination of the
critical exponents of the site-diluted Gaussian EA model is not in the
scope of this paper as the statistical quality of our data for this
model is clearly poorer than the one of the data for the bond-diluted
$\pm J$ EA model, as can be seen from Tables \ref{tab:simulation_1} and
\ref{tab:simulation_2}.

\subsection{Bond-diluted $\pm J$ EA model}

\begin{figure}[tf]
  \includegraphics[width=\figurewidth]{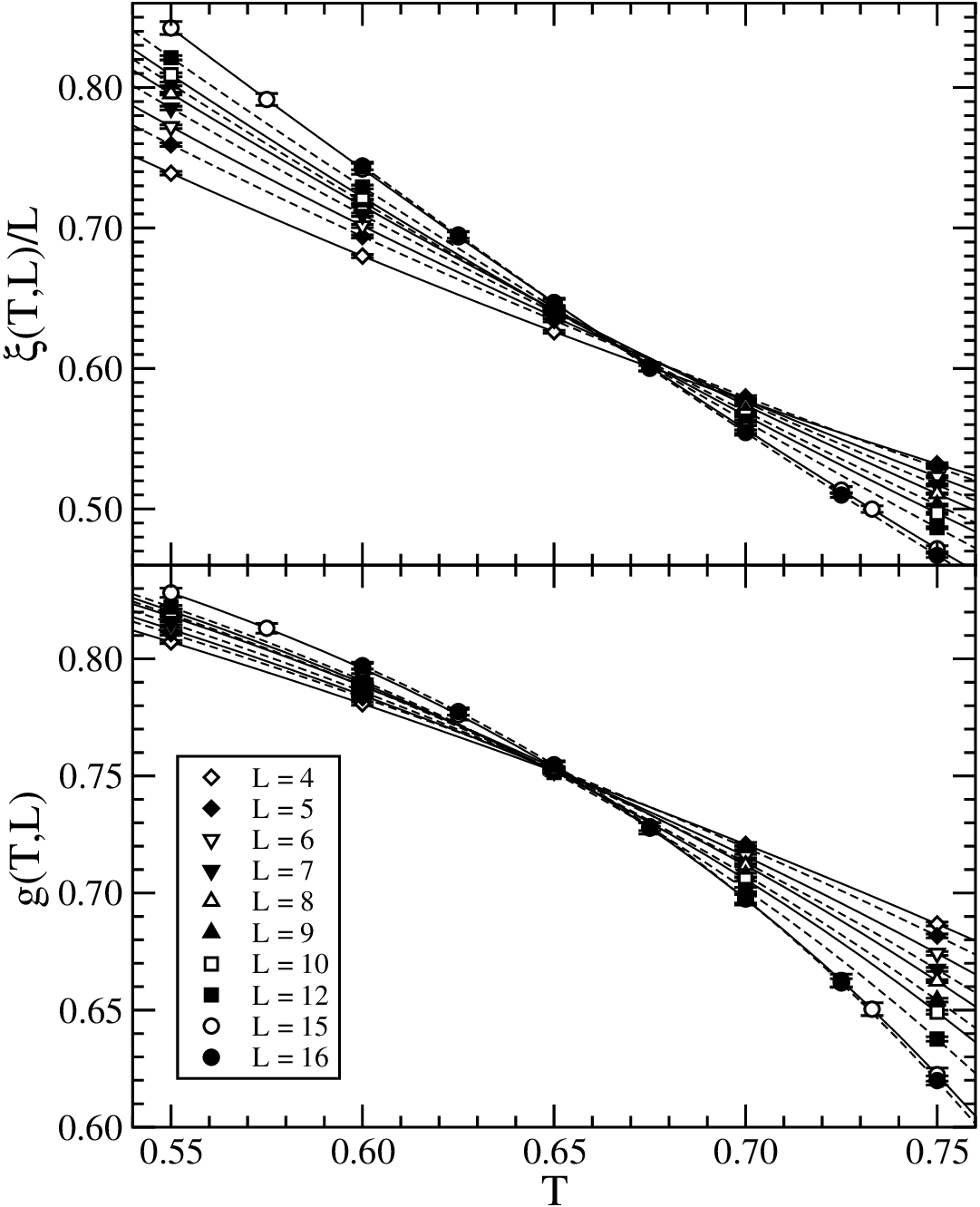}
  \caption{\label{fig:binder}The correlation length $\xi(T,L)$ measured 
    in units of the lattice size $L$ and the Binder ratio $g(T,L)$ for
    the different system sizes in function of the temperature $T$.
    Note that the intersection of the different curves is very clean
    apart from that of the smallest system with $L=4$ and define $T_c$
    with good precision. The crossings of $\xi(T,L)/L$ and $g(T,L)$
    occur very close to each other which is an indication of small
    scaling violations. In order to make the inversion of the ordering
    below and above the phase transition more evident we have
    connected the data points with cubic splines.}
\end{figure}

In Fig.~\ref{fig:binder} we show the Binder ratio defined in
Eq.~(\ref{eq:binder}) and the correlation length from
Eq.~(\ref{eq:correlation_length}) for the different system sizes used
in the simulation (see Table \ref{tab:simulation_1}).  The
intersection of the curves for different system sizes defines an
effective critical temperature that in the infinite volume limit
converges to $T_c$.  The evidence for a SG transition is very clear
and a relatively precise and robust estimate of $T_c \approx 0.66$ can
be obtained from this data.  The intersection of the curves is very
clean, especially also for the Binder ratio, where for other coupling
distributions the crossing is not very clean.\cite{ballesteros:00,
  kawashima:96} Moreover, a comparison with the results for the
Binder ratio of the high-precision study of the undiluted $\pm J$ EA
model in Ref.~\onlinecite{ballesteros:00} gives evidence that the
corrections to scaling for the bond-diluted $\pm J$ distribution used
in our study are smaller, at least for this observable. The crossing
is also very clean for the correlation length (apart from the data of
the smallest lattice $L=4$), but this is less surprising as this
quantity typically has relatively small corrections to
scaling.\cite{ballesteros:00}

\begin{figure}[tf]
  \includegraphics[width=\figurewidth]{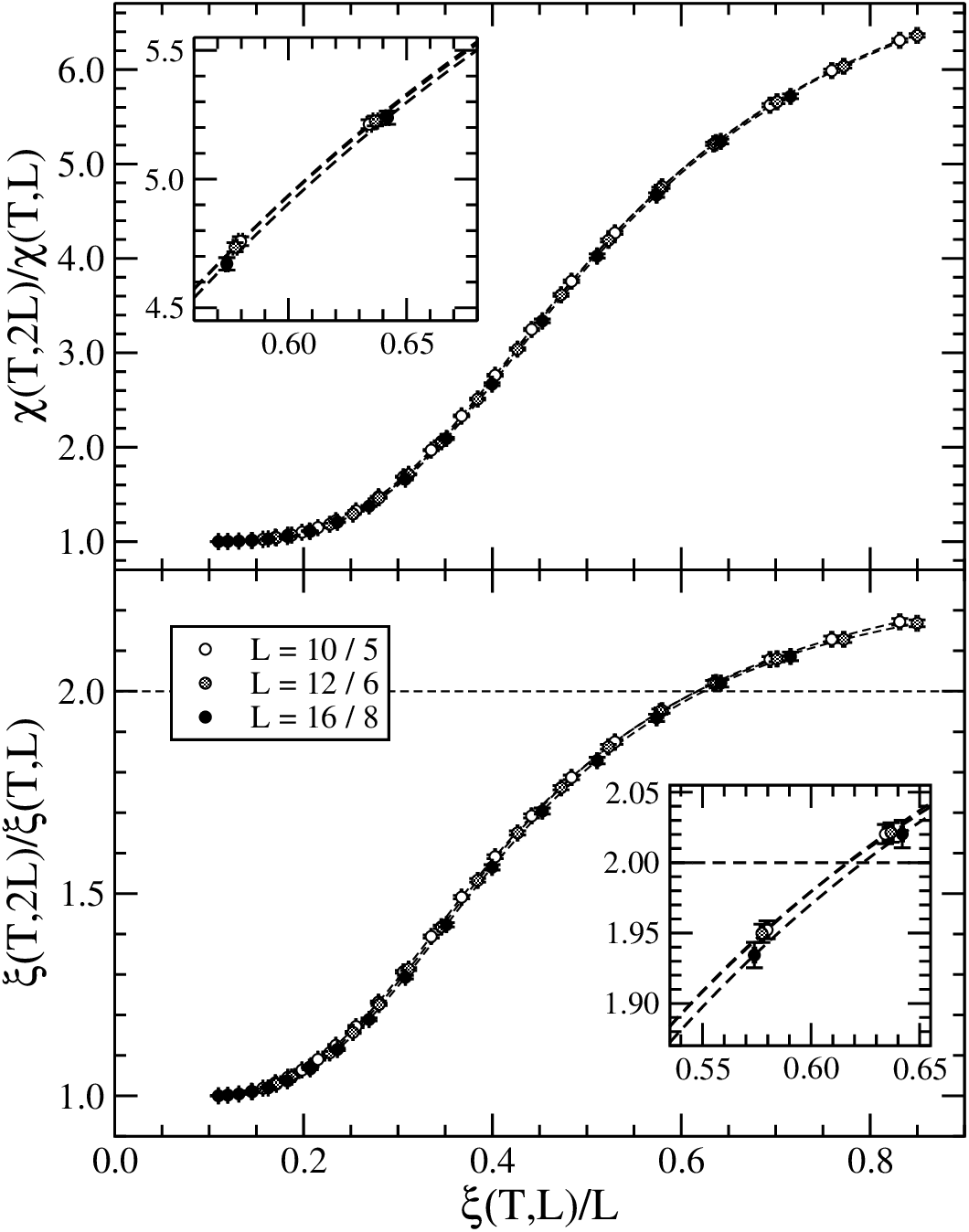}
  \caption{\label{fig:step-scaling}The finite-size scaling functions
    $F_\chi$ and $F_\xi$ for the scale factor $s=2$ from which the
    critical exponents $\nu$ and $\eta$ are obtained.  The insets show
    the same data around the effective critical point defined by
    $F_\xi = 2$. The cubic splines connecting the data points show
    that the collapse of the data for the different sizes is very
    good. This is a clear sign of small scaling violations.}
\end{figure}

In Fig.~\ref{fig:step-scaling} we show the data for the FSS functions
$F_{\xi}$ and $F_{\chi_{\rm SG}}$ defined in Eq.~(\ref{eq:FSS_2}).  The
FSS Ansatz works very well and we find very little differences between
the data for the different lattice sizes used, indicating that the
finite-size effects for these functions are already small. The data
for the pair $L=4$ and $L=8$ (not shown in
Fig.~\ref{fig:step-scaling}) on the other hand show noticeable
finite-size corrections.  In order to extrapolate the data for $\xi$
and $\chi_{\rm SG}$ to infinite volume we fit the data of $F_{\xi}$ and
$F_{\chi_{\rm SG}}$ to the ansatz $F(x) = 1 +\sum_{i=1,n} a_i \exp(-i/x)$
with $n = 5$. The fit is performed using the data of the largest
system sizes, i.e., the data of the pair $L=8$ and $L=16$. We found
only slight changes using also the data from the smaller systems and
the variations we get in the fitted quantities are well covered by the
errors given in the final results. Using the iterative procedure
described in Refs.~\onlinecite{caracciolo:95} and \onlinecite{palassini:99b} we obtain the infinite
volume data $f_{\chi_{\rm SG}}$ which allows us to extract the ratio of
the critical exponents $\gamma/\nu = 2 -\eta = 2.34 \pm 0.02$, as
asymptotically $f_{\cal{O}} \sim x^{-x_{\cal{O}}/\nu}$ for
$x\to\infty$, if $\cal{O}$ has a power-law divergence at $T_c$ with
exponent $x_{\cal{O}}$ (see Fig.~\ref{fig:scaling_comparison}).

Following Ref.~\onlinecite{palassini:99b} we fit the data extrapolated to
infinite volume to
\begin{align}\label{eq:xi_fit}
  \xi(T)&=c_\xi \, (T-T_c)^{-\nu} \, \left[ 1 + \sum_{i=0}^{n}
    a^{(i)}_\xi \, (T-T_c)^{\theta^{(i)}} \right] \\
  \label{eq:chi_fit_1} \chi_{\rm SG}(T)&=c_\chi \, (T-T_c)^{-\gamma} \,
  \left[1 + \sum_{i=0}^{n} a^{(i)}_\chi \, (T-T_c)^{\theta^{(i)}}
  \right] \\\intertext{as well as}\label{eq:chi_fit_2}
  \chi_{\rm SG}(\xi)&=b \, \xi^{2-\eta} \, \left[1 + \sum_{i=0}^{n}
    d^{(i)} \, \xi^{-\Delta^{(i)}}\right] \, ,
\end{align}
with the correction-to-scaling exponents $\theta^{(i)}$ and
$\Delta^{(i)}$. In particular we have $\theta^{(0)} = \omega \nu$ and
$\Delta^{(0)} = \omega$.  We find that including a nonleading
correction term for the fits of $\chi_{\rm SG}$ (i.e., $n = 1$ in
Eqs.~(\ref{eq:chi_fit_1}) and (\ref{eq:chi_fit_2})) reduces the
dependence on the fitting range considerably, whereas for the fits of
$\xi$ the leading correction is enough for obtaining stable and good
fits (i.e., $n = 0$ in Eq.~(\ref{eq:xi_fit})). We obtain as preferred
values
\begin{gather}
\label{eq:results_infinite} 
\nu=2.17(15), \quad \eta=-0.336(20), \quad \gamma=4.96(30), \notag\\
T_c=0.663(6)\quad \text{and}\quad \omega=0.7(3) \,. \notag
\end{gather}
Note that the values we obtain for $\gamma$, $\nu$ and $\eta$ are well
compatible with the scaling relation $\gamma = \nu(2 - \eta)$.  The
value for the correction-to-scaling exponent $\omega$ has to be taken
as an effective value as we do not separate possible analytical from
nonanalytical corrections to scaling in our fits (see
Ref.~\onlinecite{toldin:06} for a discussion).

Using on the other hand the quotient method with
Eqs.~(\ref{eq:quotient_1}) and (\ref{eq:quotient_2}) again for the
data of the pair $L=8$ and $L=16$ we obtain
\begin{gather}
  \label{eq:results_quotient} 
  \nu=2.22(15), \quad \eta=-0.349(18), \notag \\ \xi(T_c,L)/L =
  0.625(10) \,.\notag
\end{gather}
In order to fine tune the ratio $\xi/L$ on the two lattices to be
equal and in order to determine the derivative in
Eq.~(\ref{eq:quotient_2}) we use a cubic spline fit to the data. The
use of the cubic spline fits and the remaining scaling violations are
the main sources of systematic errors. We consider them, however, to
be smaller than the statistical errors in the determination of the
universal FSS functions even for the rather large number of coupling
realizations we use in this study.

\begin{figure}[tf]
  \includegraphics[width=\figurewidth]{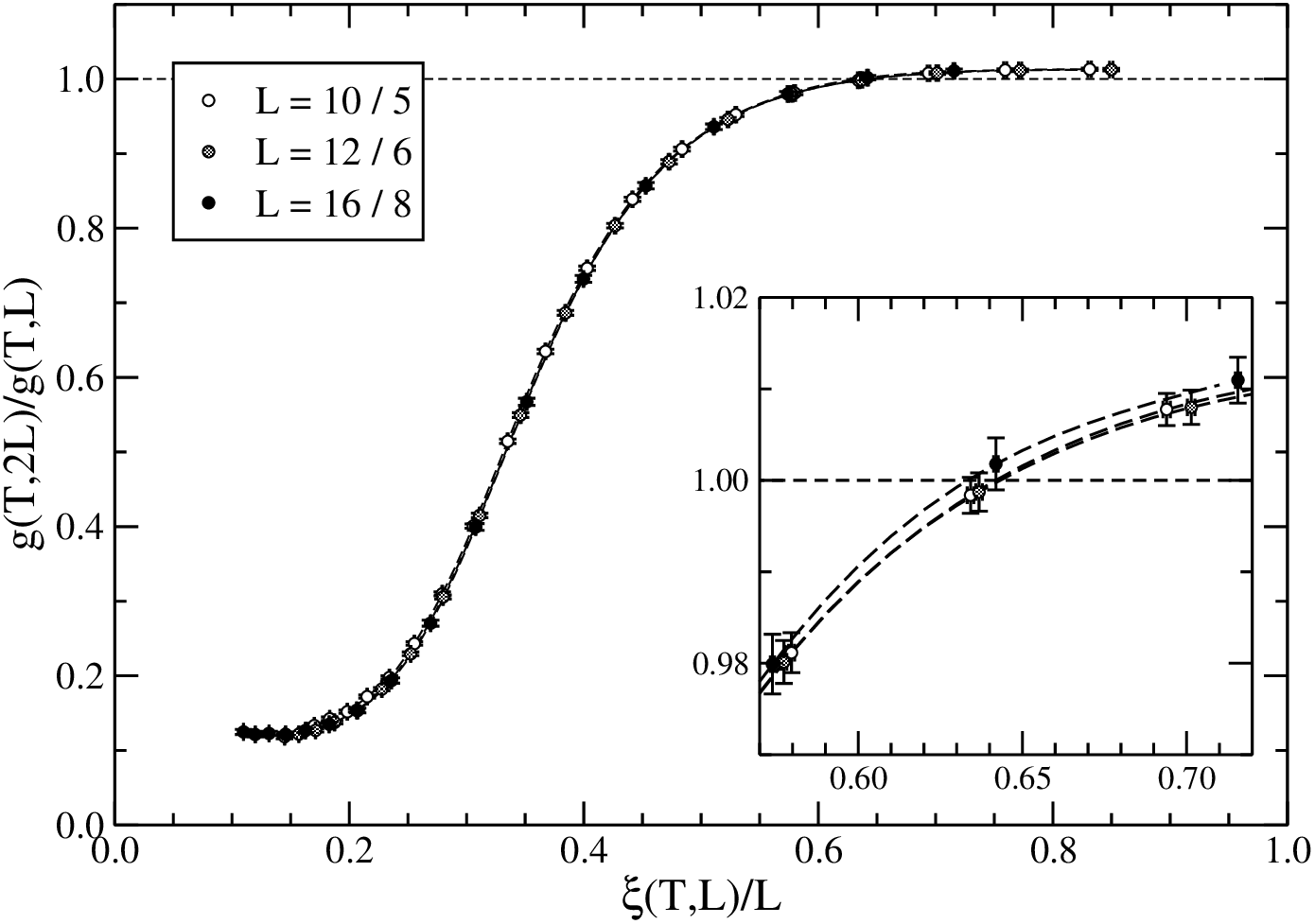}
  \caption{\label{fig:binder_scaling} The FSS function of the Binder ratio $g(T,L)$.
    The inset shows the same data around the effective critical point
    defined by $F_g = 1$. The broken lines connecting the data points
    are cubic splines making it again evident that scaling violations are
    small.}
\end{figure}

Fig.~\ref{fig:binder_scaling} shows the FSS function $F_{g}$ of the
Binder ratio $g$.  The value of $\xi(T_c^*,L)/L$ at which $F_{g} = 1$
gives an alternative possibility to define an effective critical
point. The fact that $\xi(T_c^*,L)/L$ defined from $F_{g} = 1$ (shown
in the inset of Fig.~\ref{fig:binder_scaling}) is slightly larger than
the one defined from $F_{\xi} = 2$ (see inset of
Fig.~\ref{fig:step-scaling}) indicates that there are remaining
corrections to scaling. The value of $\xi(T_c^*,L)/L$ taken from the
pair $L=8$ and $L=16$ and the condition $F_g = 1$ is $\xi(T_c^*,L)/L =
0.635(15)$.  Comparing to the corresponding value from the pair $L=8$
and $L=16$ and the condition $F_{\xi} = 2$, which is $\xi(T_c^*,L)/L =
0.625(10)$, we can conclude that these remaining corrections to
scaling are small, however.

\begin{figure}[htbf]
  \includegraphics[width=\figurewidth]{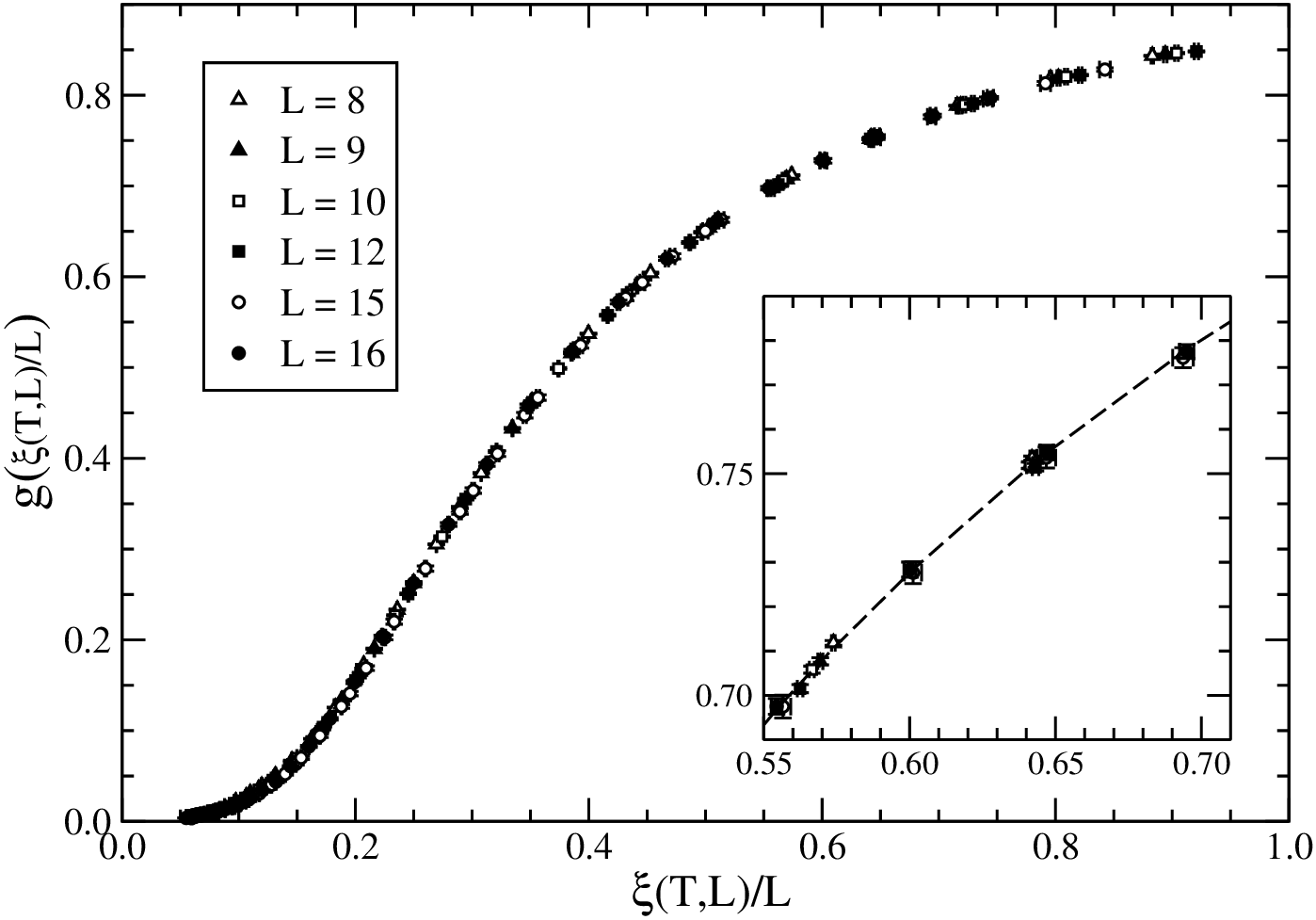}
  \caption{\label{fig:binder_xi} The universal FSS function defined by the Binder
    ratio $g(T,L)$ versus the correlation length $\xi(T,L)/L$.  Note
    that due to the very small scaling violations the plot defines a
    precise mapping between the two quantities that are typically used
    to define the critical temperature. The inset shows the same data
    around the effective critical point.  The broken line is a
    polynomial fit through all the data points (i.e., $L=8, \ldots
    ,16$) in the range $0.5\leq\xi(T,L)/L \leq 0.8$.}
\end{figure}

In Fig.~\ref{fig:binder_xi} we show the universal FSS function defined
by the Binder ratio $g(T,L)$ as a function of the correlation length
$\xi(T,L)/L$. This FSS function is particularly interesting as these
two dimensionless quantities have a rather large cross correlation.
This means that all the data for the sizes from $L=8$ to $16$ fall on
the same curve within their statistical errors. The curve provides a
one-to-one correspondence between the two quantities that are often
used to determine the critical point in SG simulations. Using the
value $\xi(T_c,L)/L = 0.625(10)$ from the condition $F_{\xi} = 2$ as
our most reliable determination of this quantity we obtain a
corresponding value of the Binder ratio at the critical point of
$g(T_c) = 0.742(7)$.

\subsection{Site-diluted Gaussian EA model}

\begin{figure}[tf]
  \includegraphics[width=\figurewidth]{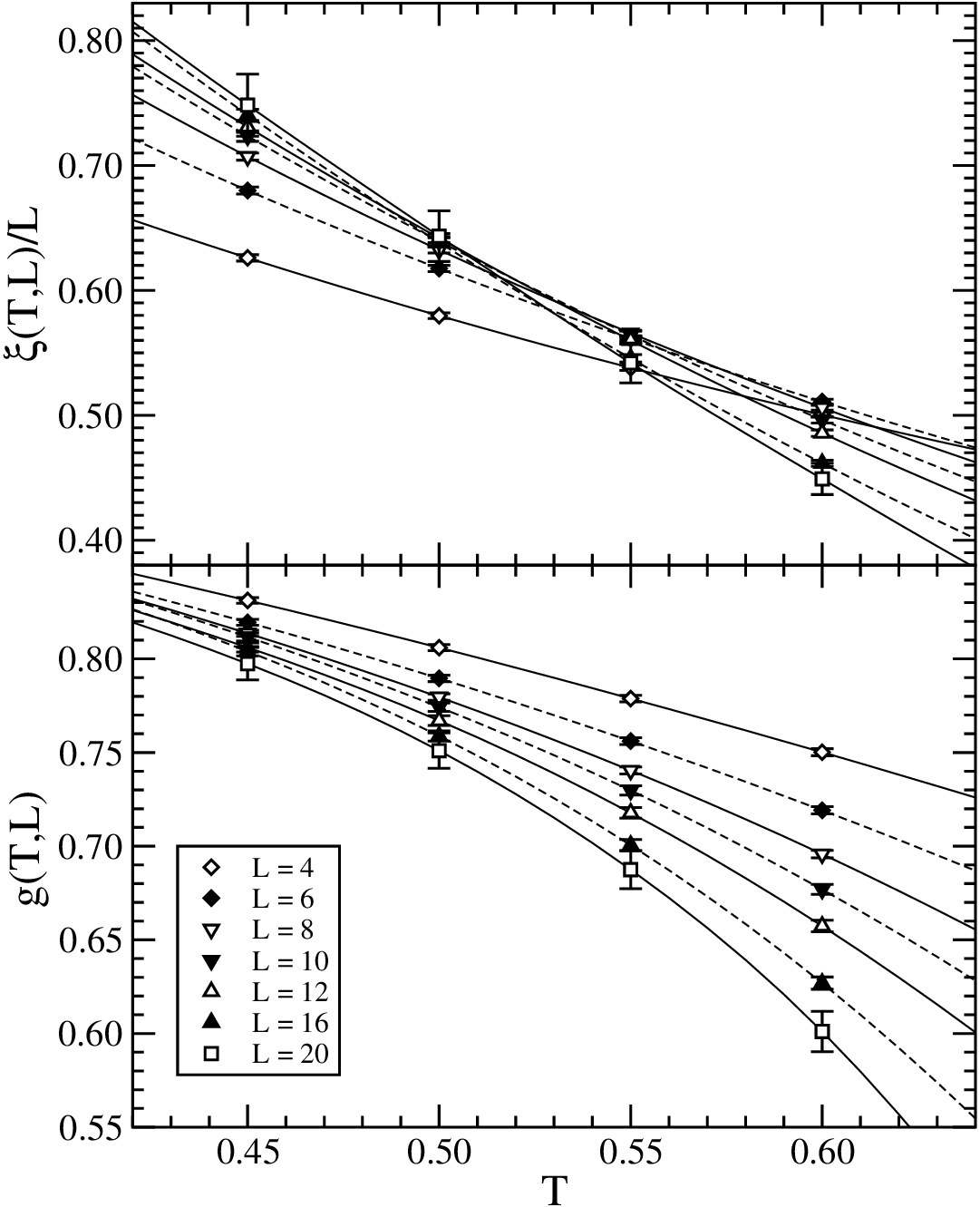}
  \caption{\label{fig:binder_gauss} The same as Fig.~\ref{fig:binder},
    but for the site-diluted Gaussian coupling distribution.  The
    correlation length shows a crossing at $T_c \approx 0.51$. The
    large shift between the crossing points for the smaller system
    sizes indicates rather large finite-size corrections. These large
    scaling corrections can be seen even better for the Binder ratio,
    as it does not even cross for the different system sizes and
    temperature range used in this simulation. The data points are
    connected by cubic splines.}
\end{figure}

In Fig.~\ref{fig:binder_gauss} we show the Binder ratio defined in
Eq.~(\ref{eq:binder}) and the correlation length from
Eq.~(\ref{eq:correlation_length}) for the different system sizes used
in the simulation (see Table \ref{tab:simulation_2}). In contrast to
the data for the bond-diluted $\pm J$ model shown in
Fig.~\ref{fig:binder} the crossing of the curves for the correlation
length is no longer very neat. Moreover, the Binder ratio does not
cross at all for the system sizes and the temperature range used in
this simulation.  With increasing system sizes the crossing of the
correlation length shifts from larger $T$ values toward $T \approx
0.51$. Although there is no crossing of the Binder ratio we consider
the data of the correlation length to be clear enough evidence for a
SG transition around $T_c \approx 0.51$.  We attribute the fact that
there is no clean intersection to the presence of very strong
corrections to scaling in this model and we will give evidence for
this view in Sec.~\ref{sec:universality}. We think that the
corrections are so large that nonleading correction terms cannot be
neglected.

\section{Universality of the FSS functions}
\label{sec:universality}

It is obviously interesting to see how the FSS functions of these two
coupling distributions having such a different level of scaling
corrections compare to each other and that is what we are going to do in
this section. But first we compare the FSS functions of the
bond-diluted $\pm J$ EA model with the ones of the undiluted $\pm J$ EA
model.

\begin{figure}[tbf]
  \includegraphics[width=0.75\figurewidth]{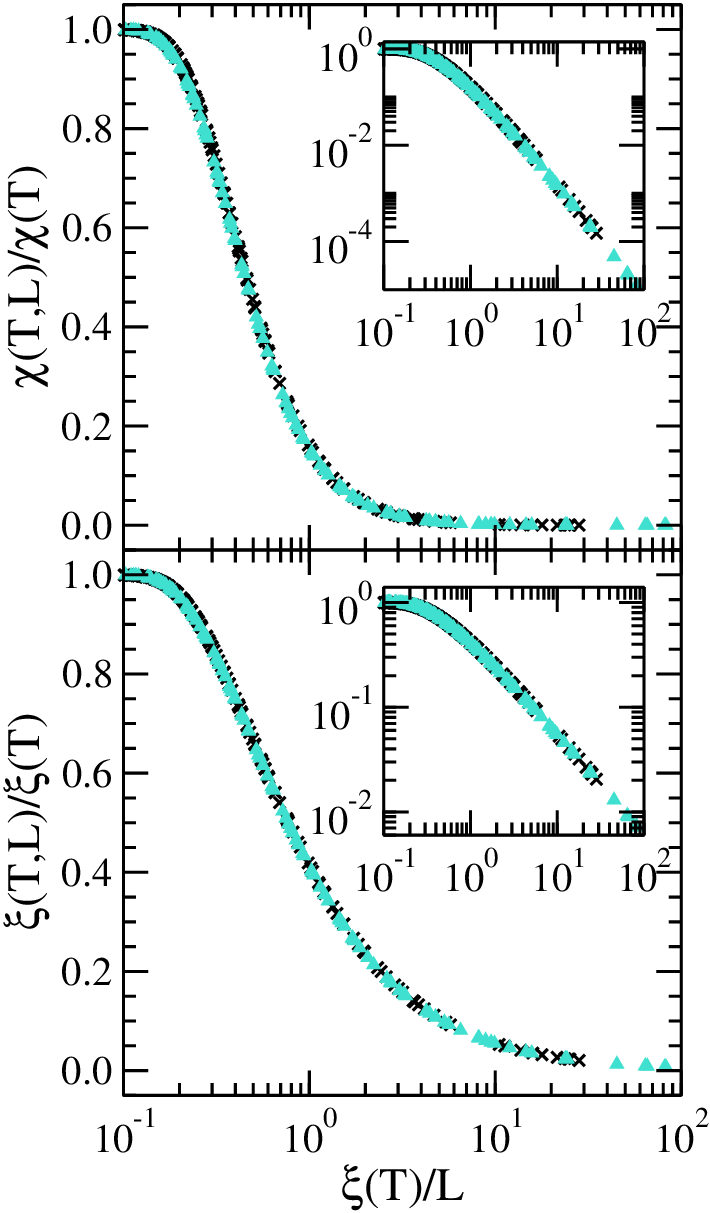}
  \caption{\label{fig:scaling_comparison} (Color online) A comparison 
    of the universal FSS functions $f_\chi$ and $f_\xi$ for the scale
    factor $s=2$ of the bond-diluted $\pm J$ EA model (circles, $L=4$
    to $L=16$) with the corresponding data for the undiluted $\pm J$ EA
    model from Palassini and Caracciolo\cite{palassini:99b}
    (triangles, $L=5$ to $L=48$).  The insets show the same data in a
    log-log plot, making evident the power-law decay at large
    $\xi(T,\infty)/L$. The excellent agreement between the two data
    sets gives strong evidence that above the critical dilution $p^*$
    the critical behavior of the 3D $\pm J$ EA model does {\it not}
    depend on bond dilution.}
\end{figure}

In Fig.~\ref{fig:scaling_comparison} we compare the data for the FSS
functions $f_{\xi}$ and $f_{\chi_{\rm SG}}$ defined in
Eq.~(\ref{eq:FSS_1}) obtained in this simulation with the data from
the 3D undiluted $\pm J$ EA model obtained by Palassini and Caracciolo
in Ref.~\onlinecite{palassini:99b}.  The agreement of the two data
sets is excellent giving very strong evidence that above the critical
dilution $p^*$ the critical behavior of the $\pm J$ EA model does {\it
  not} depend on bond dilution.

\begin{figure}[tbf]
  \includegraphics[width=\figurewidth]{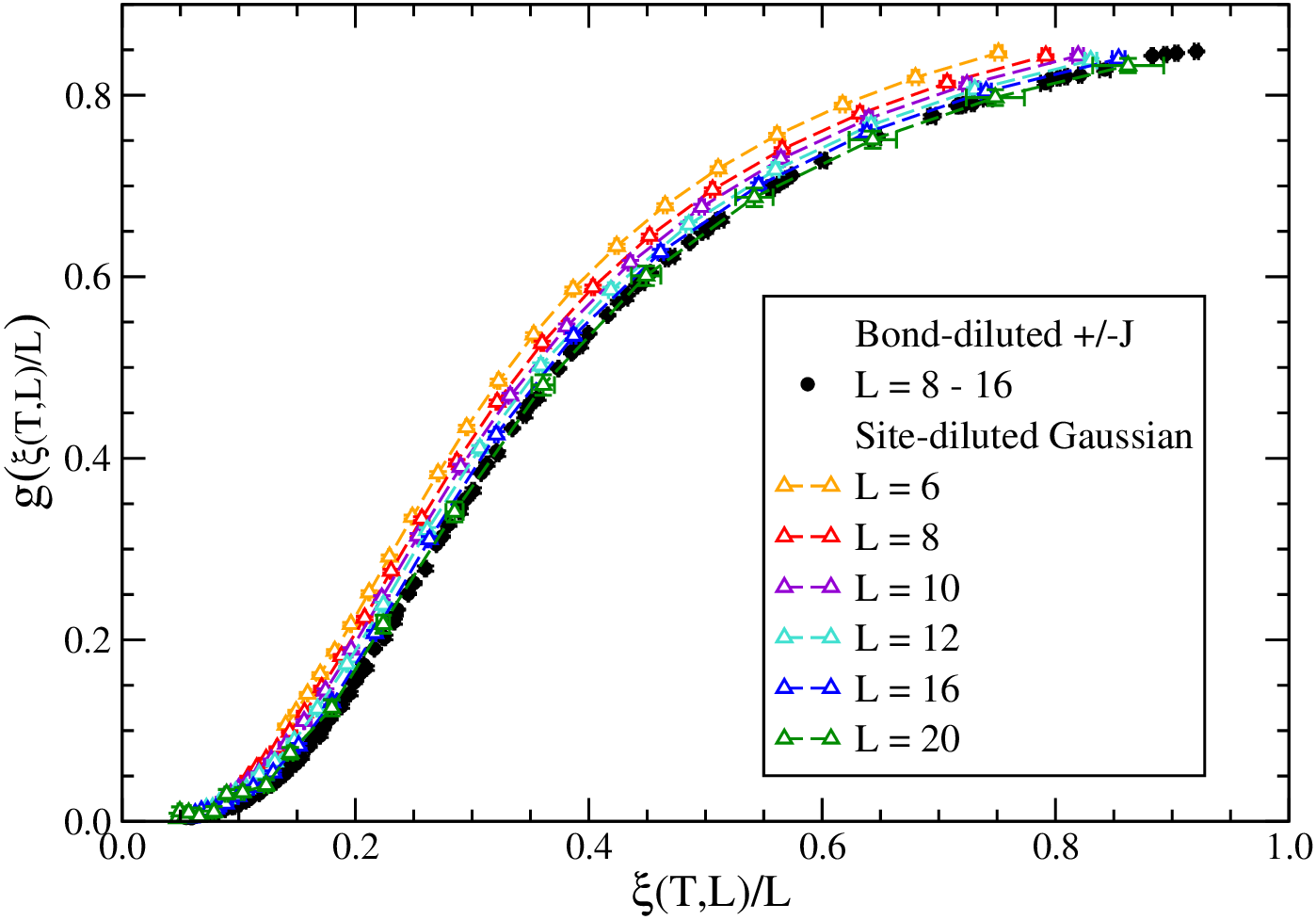}
  \caption{\label{fig:binder_xi_comparison} (Color online) A comparison 
    of the universal FSS function defined by the Binder ratio $g(T,L)$
    versus the correlation length $\xi(T,L)/L$ (see also
    Fig.~\ref{fig:binder_xi}) is given for the bond-diluted $\pm J$ and
    the site-diluted Gaussian EA model. The figure shows how the data
    of the site-diluted Gaussian model with increasing system size
    converges toward the curve defined by the data of the
    bond-diluted $\pm J$ model. The data for the site-diluted Gaussian
    EA model is connected by cubic splines to make the convergence
    more evident.}
\end{figure}

In Fig.~\ref{fig:binder_xi_comparison} we compare the data for the FSS
function defined by the Binder ratio $g(T,L)$ versus the correlation
length $\xi(T,L)/L$ for the bond-diluted $\pm J$ with the site-diluted
Gaussian coupling distribution.  In Fig.~\ref{fig:binder_xi} we showed
that this function shows very little scaling corrections for the
bond-diluted $\pm J$ EA model. Not surprisingly this is not the case
for the site-diluted Gaussian EA model. The curves for increasing
system sizes are, however, moving closer and closer to the curve
defined by the data of the bond-diluted $\pm J$ EA model. The agreement
between the curves for the largest few system sizes is very good
giving support to the claim that the two models fall into the same
universality class.

\begin{figure}[tbf]
  \includegraphics[width=\figurewidth]{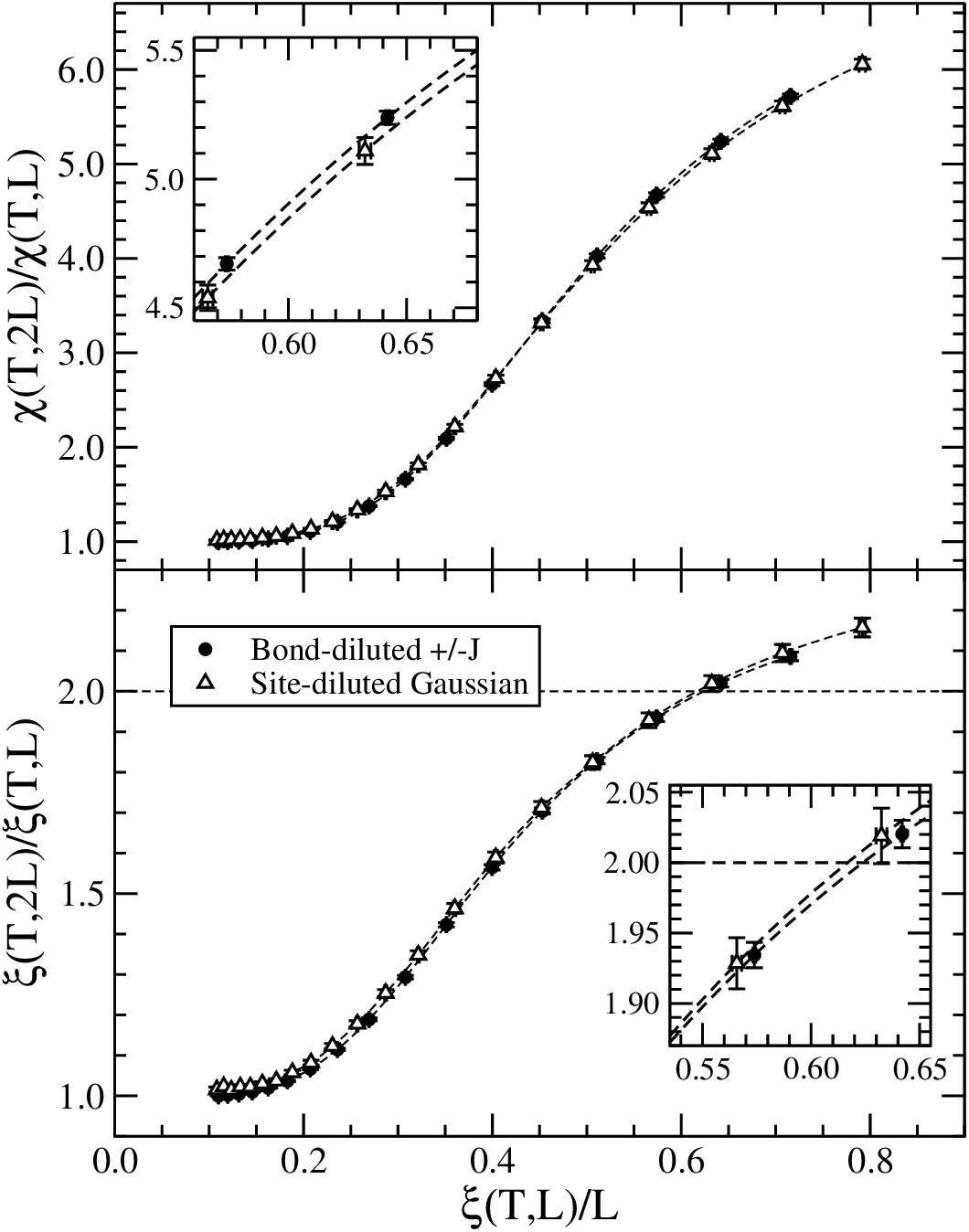}
  \caption{\label{fig:xi_chi_comparison} A comparison of the 
    finite-size scaling functions $F_\chi$ and $F_\xi$ for the scale
    factor $s=2$ is given for the bond-diluted $\pm J$ and the
    site-diluted Gaussian EA model for the pair $L=8$ and $L=16$.  The
    insets show the same data around the effective critical point
    defined by $F_\xi = 2$. From the fact that the data collapse is
    very good we can conclude that the estimates for the critical
    exponents $\nu$ and $\eta$ for the two models are the same within
    errors.}
\end{figure}

In Fig.~\ref{fig:xi_chi_comparison} we compare the data for the FSS
functions $F_{\xi}$ and $F_{\chi_{\rm SG}}$ as defined in
Eq.~(\ref{eq:FSS_2}) of the bond-diluted $\pm J$ EA model with
corresponding ones of the site-diluted Gaussian EA model for the pair
$L=8$ and $L=16$.  The agreement is again very good. Without
performing any calculation we can read off this figure that the
critical exponents $\eta$ and $\nu$ of the two models are fully
compatible with each other within errors.

\section{Discussion and conclusions}
Replica cluster MC algorithms
\cite{swendsen:86,redner:98,houdayer:01,wang:05} do not only allow for
an efficient simulation of the two-dimensional EA model, but also for
different 3D diluted models with disorder and
frustration.\cite{jorg:05} We have applied a variant of such an
algorithm to the simulation of the 3D bond-diluted EA model with
binary couplings. The universal FSS functions are determined with very
high statistical accuracy and they show little size dependence
indicating that the corresponding finite-size corrections are becoming
small when determined from a pair of systems with $L=5$ and $L=10$ or
larger. Clearly a change of the behavior of the FSS functions at
larger sizes than considered here cannot completely be excluded, but
seems rather improbable. In order to obtain good knowledge on the
magnitude of the remaining scaling corrections a more precise study
with even more statistics would be needed.  Extrapolating the data for
the correlation length $\xi$ and the SG susceptibility $\chi_{\rm SG}$ to
infinite volume we obtain the critical exponents $\nu = 2.17(15)$,
$\eta = -0.336(20)$, and the effective leading correction-to-scaling
exponent $\omega = 0.7(3)$.  The value of $\gamma = 4.96(30)$ that we
obtain is well compatible with values of $\nu$, $\eta$ and the scaling
relation $\gamma = \nu (2 - \eta)$ connecting these exponents. For the
value of the critical temperature we get $T_c = 0.663(6)$. On the
other hand using the quotient method we extract $\nu = 2.22(15)$ and
$\eta = -0.349(18)$. Other critical quantities we obtain from this
method are $\xi(T_c,L)/L = 0.625(10)$ and using the one-to-one
correspondence between the Binder ratio and the correlation length
(see Fig.~\ref{fig:binder_xi}) we also have $g(T_c) = 0.742(7)$.  We
slightly prefer the values of the critical exponents obtained by the
quotient method over those obtained by the extrapolation to infinite
volume technique because of two reasons. The first being that the fits
are clearly more robust using the quotient method and the second being
that the exponents are extracted at the apparent critical point
$T_c^*$ which allows for a more precise numerical determination of
their values.\cite{ballesteros:97,ballesteros:00} Therefore we quote
as final values for the critical exponents of the 3D bond-diluted
$\pm J$ EA model $\nu = 2.22(15)$ and $\eta = -0.349(18)$.

The comparison of the universal FSS functions $f_{\xi}$ and
$f_{\chi_{\rm SG}}$ with the ones from the 3D undiluted $\pm J$ EA model
obtained by Palassini and Caracciolo in
Ref.~\onlinecite{palassini:99b} shows perfect agreement. This is
strong evidence that bond dilution is not a relevant perturbation,
i.e., it does {\it not} change the critical behavior of the standard
3D $\pm J$ EA model in a large range of dilutions and most probably
down to the critical dilution $p^*$ at which the SG phase appears.
This conclusion is furthermore supported by the excellent agreement
between the critical values obtained in this study [$\nu = 2.22(15)$,
$\eta = -0.349(18)$] with the ones from the high-precision study in
Ref.~\onlinecite{ballesteros:00}, where $\nu = 2.15(15)$ and $\eta =
-0.337(15)$ are given.
 
At this point a remark on the difference between the critical
exponents obtained by Palassini and Caracciolo\cite{palassini:99b} is
in order, since they give $\nu=1.8(2)$ and $\eta=-0.26(4)$ using the
same infinite volume extrapolation technique used in this paper and
moreover as they have (within errors) the same universal FSS functions
$f_{\xi}$ and $f_{\chi_{\rm SG}}$ as we have obtained in this paper.  We
think that this discrepancy is due to the fact that a precise
determination of the critical exponents needs data of very high
accuracy close to the effective critical point. This was not the case
in the Palassini and Caracciolo\cite{palassini:99b} study that involved 
mainly data from the paramagnetic region and therefore they were unable 
to constrain the fits enough to provide exact and robust values for the 
critical exponents.

The additional comparison of the universal FSS functions $F_{\xi}$ and
$F_{\chi_{\rm SG}}$ of the bond-diluted $\pm J$ coupling distribution with
the corresponding ones of the site-diluted Gaussian coupling
distribution shows that claims for the nonuniversality of the 3D EA
model using dynamical MC \cite{bernardi:96, bernardi:97,
  mari:02,pleimling:05} are clearly not supported by a careful static
MC analysis. The agreement of the FSS functions is very good and
excludes large differences between the critical exponents of the two
models.  Smaller differences at the level of $10\%$ obviously cannot
be excluded completely with the given precision of this study, but
looking at the complete behavior of the FSS functions (and not only at
the critical point) the agreement seems too good to leave much space
for nonuniversal critical behavior of the 3D EA model.  This is
further supported by the agreement between the two models regarding
the FSS function defined by the Binder ratio versus the correlation
length.  In the case of the site-diluted Gaussian EA model this FSS
function clearly is showing large corrections to scaling for smaller
system sizes that, however, asymptotically vanish leading most
probably to a unique, i.e., universal limiting FSS function. We
conclude that from static MC simulations, especially from the
comparison of different FSS functions, we get strong evidence for
universal critical behavior of the SG transition in the 3D EA model.
The conclusion that the critical behavior of the 3D EA model is
universal is also reached in Ref.~\onlinecite{katzgraber:06}. There
the authors perform a high-precision MC simulation to study the static
properties of the model using mainly a Gaussian and a $\pm J$
distribution for the couplings. It has to be noted that the values of
the critical exponents they find, especially the one for $\eta$ of
the $\pm J$ distribution, is not fully consistent with the ones of the
present study as they find $\nu = 2.44(9)$ and $\eta = -0.37(5)$ for
the Gaussian as well as $\nu = 2.39(5)$ and $\eta = -0.395(17)$ for
the $\pm J$ distribution. We consider this difference to be due to the
different techniques to extract the critical exponents used in their
paper, which especially for the value of $\eta$ is rather sensitive
to nonanalytical as well as analytical corrections to scaling.  In
Ref.~\onlinecite{campbell:06} the 3D $\pm J$ EA model is studied on
very large lattices with a new technique to extract critical exponents.
The critical exponents obtained in this study are $\nu = 2.72(8)$ and
$\eta = -0.4(4)$ and differ again from the ones of our study and
also the ones of Ref.~\onlinecite{katzgraber:06}. We think
that the spread in the above results reflects the difficulties to
extract critical quantities in the 3D EA model with high precision and
in this respect we consider the comparison of the complete FSS
functions a more reliable and moreover parameter-free tool to check
for universality.

Finally, we have noted that the observables we have studied in this
paper show rather little scaling corrections for the 3D bond-diluted
$\pm J$ EA model at the given bond occupation of $p=0.45$.  This is
clearly in accordance with the findings of the high-temperature series
study in Ref.~\onlinecite{shapira:94} and the ground state defect
energies study in Ref.~\onlinecite{boettcher:04}. This important
physical aspect together with all the technical advantages of the
replica cluster MC algorithm and the possibility of multispin-coding
makes this model clearly a very interesting candidate for further
studies of the static properties of spin glasses.

\hspace{0.1cm}

\acknowledgments It is a pleasure to thank M.~Palassini,
F.~Ricci-Tersenghi, G.~Parisi, H.~G.~Katzgraber, A.~P.~Young, and
F.~Niedermayer for interesting discussions. The simulations have been
mainly performed on the workstations of the institute of theoretical
physics in Bern and in parts on the CLX cluster of the CINECA and the
clusters IDRA and PIOVRA in Roma. This work was supported by EEC's HPP
under contract No.~HPRN-CT-2002-00307 (DYGLAGEMEM) and ECC's FP6 IST
Programme under contract No.~IST-001935 (EVERGROW).

\bibliography{refs}

\begin{thebibliography}{40}
\expandafter\ifx\csname natexlab\endcsname\relax\def\natexlab#1{#1}\fi
\expandafter\ifx\csname bibnamefont\endcsname\relax
  \def\bibnamefont#1{#1}\fi
\expandafter\ifx\csname bibfnamefont\endcsname\relax
  \def\bibfnamefont#1{#1}\fi
\expandafter\ifx\csname citenamefont\endcsname\relax
  \def\citenamefont#1{#1}\fi
\expandafter\ifx\csname url\endcsname\relax
  \def\url#1{\texttt{#1}}\fi
\expandafter\ifx\csname urlprefix\endcsname\relax\def\urlprefix{URL }\fi
\providecommand{\bibinfo}[2]{#2}
\providecommand{\eprint}[2][]{\url{#2}}

\bibitem[{\citenamefont{Edwards and Anderson}(1975)}]{edwards:75}
\bibinfo{author}{\bibfnamefont{S.~F.} \bibnamefont{Edwards}} \bibnamefont{and}
  \bibinfo{author}{\bibfnamefont{P.~W.} \bibnamefont{Anderson}},
  \emph{\bibinfo{title}{Theory of spin glasses}}, \bibinfo{journal}{J. Phys. F:
  Met. Phys.} \textbf{\bibinfo{volume}{5}}, \bibinfo{pages}{965}
  (\bibinfo{year}{1975}).

\bibitem[{\citenamefont{Diep}(2005)}]{diep_cha:05}
\bibinfo{editor}{\bibfnamefont{H.~T.} \bibnamefont{Diep}}, ed.,
  \emph{\bibinfo{title}{Frustrated {S}pin {S}ystems}}
  (\bibinfo{publisher}{World {S}cientific}, \bibinfo{year}{2005}),
  chap.~\bibinfo{chapter}{9}.

\bibitem[{\citenamefont{Swendsen and Wang}(1986)}]{swendsen:86}
\bibinfo{author}{\bibfnamefont{R.~H.} \bibnamefont{Swendsen}} \bibnamefont{and}
  \bibinfo{author}{\bibfnamefont{J.-S.} \bibnamefont{Wang}},
  \emph{\bibinfo{title}{Replica {M}onte {C}arlo simulation of spin-glasses}},
  \bibinfo{journal}{Phys. Rev. Lett.} \textbf{\bibinfo{volume}{57}},
  \bibinfo{pages}{2607} (\bibinfo{year}{1986}).

\bibitem[{\citenamefont{Redner et~al.}(1998)\citenamefont{Redner, Machta, and
  Chayes}}]{redner:98}
\bibinfo{author}{\bibfnamefont{O.}~\bibnamefont{Redner}},
  \bibinfo{author}{\bibfnamefont{J.}~\bibnamefont{Machta}}, \bibnamefont{and}
  \bibinfo{author}{\bibfnamefont{L.~F.} \bibnamefont{Chayes}},
  \emph{\bibinfo{title}{Graphical representations and cluster algorithms for
  critical points with fields}}, \bibinfo{journal}{Phys. Rev. E}
  \textbf{\bibinfo{volume}{58}}, \bibinfo{pages}{2749} (\bibinfo{year}{1998}).

\bibitem[{\citenamefont{Houdayer}(2001)}]{houdayer:01}
\bibinfo{author}{\bibfnamefont{J.}~\bibnamefont{Houdayer}},
  \emph{\bibinfo{title}{A cluster {M}onte {C}arlo algorithm for 2-dimensional
  spin glasses}}, \bibinfo{journal}{Eur. Phys. J. B.}
  \textbf{\bibinfo{volume}{22}}, \bibinfo{pages}{479} (\bibinfo{year}{2001}).

\bibitem[{\citenamefont{Wang and Swendsen}(2005)}]{wang:05}
\bibinfo{author}{\bibfnamefont{J.-S.} \bibnamefont{Wang}} \bibnamefont{and}
  \bibinfo{author}{\bibfnamefont{R.~H.} \bibnamefont{Swendsen}},
  \emph{\bibinfo{title}{Replica {M}onte {C}arlo simulation ({R}evisited)}},
  \bibinfo{journal}{Prog. Theor. Phys. Suppl.} \textbf{\bibinfo{volume}{157}},
  \bibinfo{pages}{317} (\bibinfo{year}{2005}).

\bibitem[{\citenamefont{J\"{o}rg}(2005)}]{jorg:05}
\bibinfo{author}{\bibfnamefont{T.}~\bibnamefont{J\"{o}rg}},
  \emph{\bibinfo{title}{Cluster {M}onte {C}arlo algorithms for diluted spin
  glasses}}, \bibinfo{journal}{Prog. Theor. Phys. Suppl.}
  \textbf{\bibinfo{volume}{157}}, \bibinfo{pages}{349} (\bibinfo{year}{2005}).

\bibitem[{\citenamefont{Shapira et~al.}(1994)\citenamefont{Shapira, Klein,
  Adler, Aharony, and Harris}}]{shapira:94}
\bibinfo{author}{\bibfnamefont{S.}~\bibnamefont{Shapira}},
  \bibinfo{author}{\bibfnamefont{L.}~\bibnamefont{Klein}},
  \bibinfo{author}{\bibfnamefont{J.}~\bibnamefont{Adler}},
  \bibinfo{author}{\bibfnamefont{A.}~\bibnamefont{Aharony}}, \bibnamefont{and}
  \bibinfo{author}{\bibfnamefont{A.~B.} \bibnamefont{Harris}},
  \emph{\bibinfo{title}{Phase diagram of the dilute {I}sing spin glass in
  general spatial dimension}}, \bibinfo{journal}{Phys. Rev. B}
  \textbf{\bibinfo{volume}{49}}, \bibinfo{pages}{8830} (\bibinfo{year}{1994}).

\bibitem[{\citenamefont{Boettcher}(2004)}]{boettcher:04}
\bibinfo{author}{\bibfnamefont{S.}~\bibnamefont{Boettcher}},
  \emph{\bibinfo{title}{Stiffness exponents for lattice spin glasses in
  dimensions d=3,...,6}}, \bibinfo{journal}{Eur. Phys. J. B}
  \textbf{\bibinfo{volume}{38}}, \bibinfo{pages}{83} (\bibinfo{year}{2004}).

\bibitem[{\citenamefont{Ballesteros et~al.}(2000)\citenamefont{Ballesteros,
  Cruz, Fern\'{a}ndez, Mart\'{i}n-Mayor, Pech, Ruiz-Lorenzo, Taranc\'{o}n,
  T\'{e}llez, Ullod, and Ungil}}]{ballesteros:00}
\bibinfo{author}{\bibfnamefont{H.~G.} \bibnamefont{Ballesteros}},
  \bibinfo{author}{\bibfnamefont{A.}~\bibnamefont{Cruz}},
  \bibinfo{author}{\bibfnamefont{L.~A.} \bibnamefont{Fern\'{a}ndez}},
  \bibinfo{author}{\bibfnamefont{V.}~\bibnamefont{Mart\'{i}n-Mayor}},
  \bibinfo{author}{\bibfnamefont{J.}~\bibnamefont{Pech}},
  \bibinfo{author}{\bibfnamefont{J.~J.} \bibnamefont{Ruiz-Lorenzo}},
  \bibinfo{author}{\bibfnamefont{A.}~\bibnamefont{Taranc\'{o}n}},
  \bibinfo{author}{\bibfnamefont{P.}~\bibnamefont{T\'{e}llez}},
  \bibinfo{author}{\bibfnamefont{C.~L.} \bibnamefont{Ullod}}, \bibnamefont{and}
  \bibinfo{author}{\bibfnamefont{C.}~\bibnamefont{Ungil}},
  \emph{\bibinfo{title}{Critical behavior of the three-dimensional {I}sing spin
  glass}}, \bibinfo{journal}{Phys. Rev. B} \textbf{\bibinfo{volume}{62}},
  \bibinfo{pages}{14237} (\bibinfo{year}{2000}).

\bibitem[{\citenamefont{Palassini and Caracciolo}(1999)}]{palassini:99b}
\bibinfo{author}{\bibfnamefont{M.}~\bibnamefont{Palassini}} \bibnamefont{and}
  \bibinfo{author}{\bibfnamefont{S.}~\bibnamefont{Caracciolo}},
  \emph{\bibinfo{title}{{U}niversal {F}inite-{S}ize {S}caling {F}unctions in
  the 3{D} {I}sing {S}pin {G}lass}}, \bibinfo{journal}{Phys. Rev. Lett.}
  \textbf{\bibinfo{volume}{82}}, \bibinfo{pages}{5128} (\bibinfo{year}{1999}).

\bibitem[{\citenamefont{Bernardi et~al.}(1996)\citenamefont{Bernardi, Prakash,
  and Campbell}}]{bernardi:96}
\bibinfo{author}{\bibfnamefont{L.~W.} \bibnamefont{Bernardi}},
  \bibinfo{author}{\bibfnamefont{S.}~\bibnamefont{Prakash}}, \bibnamefont{and}
  \bibinfo{author}{\bibfnamefont{I.~A.} \bibnamefont{Campbell}},
  \emph{\bibinfo{title}{{O}rdering {T}emperatures and {C}ritical {E}xponents in
  {I}sing {S}pin {G}lasses}}, \bibinfo{journal}{Phys. Rev. Lett.}
  \textbf{\bibinfo{volume}{77}}, \bibinfo{pages}{2798} (\bibinfo{year}{1996}).

\bibitem[{\citenamefont{Bernardi and Campbell}(1997)}]{bernardi:97}
\bibinfo{author}{\bibfnamefont{L.~W.} \bibnamefont{Bernardi}} \bibnamefont{and}
  \bibinfo{author}{\bibfnamefont{I.~A.} \bibnamefont{Campbell}},
  \emph{\bibinfo{title}{Critical exponents in {I}sing spin glasses}},
  \bibinfo{journal}{Phys. Rev. B} \textbf{\bibinfo{volume}{56}},
  \bibinfo{pages}{5271} (\bibinfo{year}{1997}).

\bibitem[{\citenamefont{Mari and Campbell}(2002)}]{mari:02}
\bibinfo{author}{\bibfnamefont{P.~O.} \bibnamefont{Mari}} \bibnamefont{and}
  \bibinfo{author}{\bibfnamefont{I.~A.} \bibnamefont{Campbell}},
  \emph{\bibinfo{title}{The ordering temperature and critical exponents of the
  bimodal {I}sing spin glass in dimension three}}, \bibinfo{journal}{Phys. Rev.
  B} \textbf{\bibinfo{volume}{65}}, \bibinfo{pages}{184409}
  (\bibinfo{year}{2002}).

\bibitem[{\citenamefont{Daboul et~al.}(2004)\citenamefont{Daboul, Chang, and
  Aharony}}]{daboul:04}
\bibinfo{author}{\bibfnamefont{D.}~\bibnamefont{Daboul}},
  \bibinfo{author}{\bibfnamefont{I.}~\bibnamefont{Chang}}, \bibnamefont{and}
  \bibinfo{author}{\bibfnamefont{A.}~\bibnamefont{Aharony}},
  \emph{\bibinfo{title}{Test of universality in the {I}sing spin glass using
  high temperature graph expansion}}, \bibinfo{journal}{Eur. Phys. J. B}
  \textbf{\bibinfo{volume}{41}}, \bibinfo{pages}{231} (\bibinfo{year}{2004}).

\bibitem[{\citenamefont{Pleimling and Campbell}(2005)}]{pleimling:05}
\bibinfo{author}{\bibfnamefont{M.}~\bibnamefont{Pleimling}} \bibnamefont{and}
  \bibinfo{author}{\bibfnamefont{I.~A.} \bibnamefont{Campbell}},
  \emph{\bibinfo{title}{{Dynamic critical behavior in Ising spin glasses}}},
  \bibinfo{journal}{Phys. Rev. B} \textbf{\bibinfo{volume}{72}},
  \bibinfo{pages}{184429} (\bibinfo{year}{2005}).

\bibitem[{\citenamefont{Katzgraber et~al.}(2006)\citenamefont{Katzgraber,
  K\"orner, and Young}}]{katzgraber:06}
\bibinfo{author}{\bibfnamefont{H.~G.} \bibnamefont{Katzgraber}},
  \bibinfo{author}{\bibfnamefont{M.}~\bibnamefont{K\"orner}}, \bibnamefont{and}
  \bibinfo{author}{\bibfnamefont{A.~P.} \bibnamefont{Young}},
  \emph{\bibinfo{title}{{Universality in three-dimensional Ising spin glasses:
  A Monte Carlo study}}}, \bibinfo{journal}{Phys. Rev. B}
  \textbf{\bibinfo{volume}{73}}, \bibinfo{pages}{224432}
  (\bibinfo{year}{2006}).

\bibitem[{\citenamefont{Parisen~Toldin
  et~al.}(2006)\citenamefont{Parisen~Toldin, Pelissetto, and
  Vicari}}]{toldin:06}
\bibinfo{author}{\bibfnamefont{F.}~\bibnamefont{Parisen~Toldin}},
  \bibinfo{author}{\bibfnamefont{A.}~\bibnamefont{Pelissetto}},
  \bibnamefont{and} \bibinfo{author}{\bibfnamefont{E.}~\bibnamefont{Vicari}},
  \emph{\bibinfo{title}{Critical behaviour of the random-anisotropy model in
  the strong-anisotropy limit}}, \bibinfo{journal}{J. Stat. Mech.}
  \textbf{\bibinfo{volume}{2006}}, \bibinfo{pages}{P06002}
  (\bibinfo{year}{2006}).

\bibitem[{\citenamefont{Ballesteros et~al.}(1997)\citenamefont{Ballesteros,
  Fern\'{a}ndez, Mart\'{i}n-Mayor, and Mu\~{n}oz Sudupe}}]{ballesteros:97}
\bibinfo{author}{\bibfnamefont{H.~G.} \bibnamefont{Ballesteros}},
  \bibinfo{author}{\bibfnamefont{L.~A.} \bibnamefont{Fern\'{a}ndez}},
  \bibinfo{author}{\bibfnamefont{V.}~\bibnamefont{Mart\'{i}n-Mayor}},
  \bibnamefont{and} \bibinfo{author}{\bibfnamefont{A.}~\bibnamefont{Mu\~{n}oz
  Sudupe}}, \emph{\bibinfo{title}{Critical properties of the
  {A}ntiferromagnetic ${R}{P}^2$ model in three dimensions}},
  \bibinfo{journal}{Nucl. Phys. B} \textbf{\bibinfo{volume}{483}},
  \bibinfo{pages}{707} (\bibinfo{year}{1997}).

\bibitem[{\citenamefont{Caracciolo et~al.}(1995)\citenamefont{Caracciolo,
  Edwards, Ferreira, Pelissetto, and Sokal}}]{caracciolo:95}
\bibinfo{author}{\bibfnamefont{S.}~\bibnamefont{Caracciolo}},
  \bibinfo{author}{\bibfnamefont{R.~G.} \bibnamefont{Edwards}},
  \bibinfo{author}{\bibfnamefont{S.~J.} \bibnamefont{Ferreira}},
  \bibinfo{author}{\bibfnamefont{A.}~\bibnamefont{Pelissetto}},
  \bibnamefont{and} \bibinfo{author}{\bibfnamefont{A.~D.} \bibnamefont{Sokal}},
  \emph{\bibinfo{title}{Extrapolating {M}onte {C}arlo {S}imulations to
  {I}nfinite {V}olume: {F}inite-{S}ize {S}caling at $\xi/{L} \gg 1$}},
  \bibinfo{journal}{Phys. Rev. Lett.} \textbf{\bibinfo{volume}{74}},
  \bibinfo{pages}{2969} (\bibinfo{year}{1995}).

\bibitem[{\citenamefont{Harris}(1974)}]{harris:74}
\bibinfo{author}{\bibfnamefont{A.~B.} \bibnamefont{Harris}},
  \emph{\bibinfo{title}{Effect of random defects on the critical behaviour of
  {I}sing models}}, \bibinfo{journal}{J. Phys. C} \textbf{\bibinfo{volume}{7}},
  \bibinfo{pages}{1671} (\bibinfo{year}{1974}).

\bibitem[{\citenamefont{Parisi et~al.}(1999)\citenamefont{Parisi,
  Ricci-Tersenghi, and Ruiz-Lorenzo}}]{parisi:99x}
\bibinfo{author}{\bibfnamefont{G.}~\bibnamefont{Parisi}},
  \bibinfo{author}{\bibfnamefont{F.}~\bibnamefont{Ricci-Tersenghi}},
  \bibnamefont{and} \bibinfo{author}{\bibfnamefont{J.~J.}
  \bibnamefont{Ruiz-Lorenzo}}, \emph{\bibinfo{title}{Universality in the
  off-equilibrium critical dynamics of the $3d$ diluted ising model}},
  \bibinfo{journal}{Phys. Rev. E} \textbf{\bibinfo{volume}{60}},
  \bibinfo{pages}{5198} (\bibinfo{year}{1999}).

\bibitem[{\citenamefont{Jim\'{e}nez and J\"{o}rg}()}]{jimenez:06}
\bibinfo{author}{\bibfnamefont{S.}~\bibnamefont{Jim\'{e}nez}} \bibnamefont{and}
  \bibinfo{author}{\bibfnamefont{T.}~\bibnamefont{J\"{o}rg}}, \bibinfo{note}{in
  preparation}.

\bibitem[{\citenamefont{Bray and Feng}(1987)}]{brayfeng:87}
\bibinfo{author}{\bibfnamefont{A.~J.} \bibnamefont{Bray}} \bibnamefont{and}
  \bibinfo{author}{\bibfnamefont{S.}~\bibnamefont{Feng}},
  \emph{\bibinfo{title}{Percolation of order in frustrated systems: {T}he
  dilute {J} spin glass}}, \bibinfo{journal}{Phys. Rev. B}
  \textbf{\bibinfo{volume}{36}}, \bibinfo{pages}{8456} (\bibinfo{year}{1987}).

\bibitem[{\citenamefont{Fisher}(1972)}]{fisher:72}
\bibinfo{author}{\bibfnamefont{M.~E.} \bibnamefont{Fisher}}, in
  \emph{\bibinfo{booktitle}{Proceedings of the 51st {E}nrico {F}ermi {S}ummer
  {S}chool}}, edited by \bibinfo{editor}{\bibfnamefont{M.~S.}
  \bibnamefont{Green}} (\bibinfo{publisher}{Academic {P}ress},
  \bibinfo{address}{New York}, \bibinfo{year}{1972}).

\bibitem[{\citenamefont{Privman}(1990)}]{privman:90}
\bibinfo{editor}{\bibfnamefont{V.}~\bibnamefont{Privman}}, ed.,
  \emph{\bibinfo{title}{Finite {S}ize {S}caling and {N}umerical {S}imulation of
  {S}tatistical {S}ystems}} (\bibinfo{publisher}{World Scientific},
  \bibinfo{address}{Singapore}, \bibinfo{year}{1990}).

\bibitem[{\citenamefont{Swendsen and Wang}(1987)}]{swendsen:87}
\bibinfo{author}{\bibfnamefont{R.~H.} \bibnamefont{Swendsen}} \bibnamefont{and}
  \bibinfo{author}{\bibfnamefont{J.-S.} \bibnamefont{Wang}},
  \emph{\bibinfo{title}{Nonuniversal critical dynamics in {M}onte {C}arlo
  simulations}}, \bibinfo{journal}{Phys. Rev. Lett.}
  \textbf{\bibinfo{volume}{58}}, \bibinfo{pages}{86} (\bibinfo{year}{1987}).

\bibitem[{\citenamefont{Ballesteros et~al.}(1998)\citenamefont{Ballesteros,
  Fern\'{a}ndez, Mart\'{i}n-Mayor, Mu\~{n}oz Sudupe, Parisi, and
  Ruiz-Lorenzo}}]{ballesteros:98}
\bibinfo{author}{\bibfnamefont{H.~G.} \bibnamefont{Ballesteros}},
  \bibinfo{author}{\bibfnamefont{L.~A.} \bibnamefont{Fern\'{a}ndez}},
  \bibinfo{author}{\bibfnamefont{V.}~\bibnamefont{Mart\'{i}n-Mayor}},
  \bibinfo{author}{\bibfnamefont{A.}~\bibnamefont{Mu\~{n}oz Sudupe}},
  \bibinfo{author}{\bibfnamefont{G.}~\bibnamefont{Parisi}}, \bibnamefont{and}
  \bibinfo{author}{\bibfnamefont{J.~J.} \bibnamefont{Ruiz-Lorenzo}},
  \emph{\bibinfo{title}{Critical exponents of the three-dimensional diluted
  {I}sing model}}, \bibinfo{journal}{Phys. Rev. B}
  \textbf{\bibinfo{volume}{58}}, \bibinfo{pages}{2740} (\bibinfo{year}{1998}).

\bibitem[{\citenamefont{Hukushima and Nemoto}(1996)}]{hukushima:96}
\bibinfo{author}{\bibfnamefont{K.}~\bibnamefont{Hukushima}} \bibnamefont{and}
  \bibinfo{author}{\bibfnamefont{K.}~\bibnamefont{Nemoto}},
  \emph{\bibinfo{title}{Exchange {M}onte {C}arlo method and application to spin
  glass simulations}}, \bibinfo{journal}{J. Phys. Soc. Jpn.}
  \textbf{\bibinfo{volume}{65}}, \bibinfo{pages}{1604} (\bibinfo{year}{1996}).

\bibitem[{\citenamefont{Viana and Bray}(1985)}]{viana:85}
\bibinfo{author}{\bibfnamefont{L.}~\bibnamefont{Viana}} \bibnamefont{and}
  \bibinfo{author}{\bibfnamefont{A.~J.} \bibnamefont{Bray}},
  \emph{\bibinfo{title}{Phase diagrams for dilute spin glasses}},
  \bibinfo{journal}{J. Phys. C} \textbf{\bibinfo{volume}{18}},
  \bibinfo{pages}{3037} (\bibinfo{year}{1985}).

\bibitem[{\citenamefont{Persky et~al.}(1996)\citenamefont{Persky, Kanter, and
  Solomon}}]{persky:96}
\bibinfo{author}{\bibfnamefont{N.}~\bibnamefont{Persky}},
  \bibinfo{author}{\bibfnamefont{I.}~\bibnamefont{Kanter}}, \bibnamefont{and}
  \bibinfo{author}{\bibfnamefont{S.}~\bibnamefont{Solomon}},
  \emph{\bibinfo{title}{Cluster dynamics for randomly frustrated systems with
  finite connectivity}}, \bibinfo{journal}{Phys. Rev. E}
  \textbf{\bibinfo{volume}{53}}, \bibinfo{pages}{1212} (\bibinfo{year}{1996}).

\bibitem[{\citenamefont{Niedermayer}(1988)}]{niedermayer:88}
\bibinfo{author}{\bibfnamefont{F.}~\bibnamefont{Niedermayer}},
  \emph{\bibinfo{title}{General {C}luster {U}pdating {M}ethod for {M}onte
  {C}arlo {S}imulations}}, \bibinfo{journal}{Phys. Rev. Lett.}
  \textbf{\bibinfo{volume}{61}}, \bibinfo{pages}{2026} (\bibinfo{year}{1988}).

\bibitem[{\citenamefont{Hoshen and Kopelman}(1976)}]{hoshen:76}
\bibinfo{author}{\bibfnamefont{J.}~\bibnamefont{Hoshen}} \bibnamefont{and}
  \bibinfo{author}{\bibfnamefont{R.}~\bibnamefont{Kopelman}},
  \emph{\bibinfo{title}{Percolation and cluster distribution. {I}. {C}luster
  multiple labeling technique and critical concentration algorithm}},
  \bibinfo{journal}{Phys. Rev. B} \textbf{\bibinfo{volume}{14}},
  \bibinfo{pages}{3438} (\bibinfo{year}{1976}).

\bibitem[{\citenamefont{Cooper et~al.}(1982)\citenamefont{Cooper, Freedman, and
  Preston}}]{cooper:82}
\bibinfo{author}{\bibfnamefont{F.}~\bibnamefont{Cooper}},
  \bibinfo{author}{\bibfnamefont{B.}~\bibnamefont{Freedman}}, \bibnamefont{and}
  \bibinfo{author}{\bibfnamefont{D.}~\bibnamefont{Preston}},
  \emph{\bibinfo{title}{Solving $\phi^4_{1,2}$ theory with {M}onte {C}arlo}},
  \bibinfo{journal}{Nucl. Phys. B} \textbf{\bibinfo{volume}{210}},
  \bibinfo{pages}{210} (\bibinfo{year}{1982}).

\bibitem[{\citenamefont{{Kim}}(1994)}]{kim:94}
\bibinfo{author}{\bibfnamefont{J.~K.} \bibnamefont{{Kim}}},
  \emph{\bibinfo{title}{{Asymptotic scaling of the mass gap in the
  two-dimensional ${O}(3)$ nonlinear $sigma$ model: {A} numerical study}}},
  \bibinfo{journal}{Phys. Rev. D} \textbf{\bibinfo{volume}{50}},
  \bibinfo{pages}{4663} (\bibinfo{year}{1994}).

\bibitem[{\citenamefont{Binder}(1981{\natexlab{a}})}]{binder:81}
\bibinfo{author}{\bibfnamefont{K.}~\bibnamefont{Binder}},
  \emph{\bibinfo{title}{Critical properties from {M}onte {C}arlo coarse
  graining and renormalization}}, \bibinfo{journal}{Phys. Rev. Lett.}
  \textbf{\bibinfo{volume}{47}}, \bibinfo{pages}{693}
  (\bibinfo{year}{1981}{\natexlab{a}}).

\bibitem[{\citenamefont{Binder}(1981{\natexlab{b}})}]{binder:81b}
\bibinfo{author}{\bibfnamefont{K.}~\bibnamefont{Binder}},
  \emph{\bibinfo{title}{Finite size scaling analysis of {I}sing model block
  distribution functions}}, \bibinfo{journal}{Z. Phys. B}
  \textbf{\bibinfo{volume}{43}}, \bibinfo{pages}{119}
  (\bibinfo{year}{1981}{\natexlab{b}}).

\bibitem[{\citenamefont{Efron}(1982)}]{efron:82}
\bibinfo{author}{\bibfnamefont{B.}~\bibnamefont{Efron}},
  \emph{\bibinfo{title}{The {J}ackknife, the {B}ootstrap, and {O}ther
  {R}esampling {P}lans}} (\bibinfo{publisher}{SIAM}, \bibinfo{year}{1982}).

\bibitem[{\citenamefont{Kawashima and Young}(1996)}]{kawashima:96}
\bibinfo{author}{\bibfnamefont{N.}~\bibnamefont{Kawashima}} \bibnamefont{and}
  \bibinfo{author}{\bibfnamefont{A.~P.} \bibnamefont{Young}},
  \emph{\bibinfo{title}{Phase transition in the three-dimensional $\pm {J}$
  {I}sing spin glass}}, \bibinfo{journal}{Phys. Rev. B}
  \textbf{\bibinfo{volume}{53}}, \bibinfo{pages}{R484} (\bibinfo{year}{1996}).

\bibitem[{\citenamefont{Campbell et~al.}(2006)\citenamefont{Campbell,
  Hukushima, and Takayama}}]{campbell:06}
\bibinfo{author}{\bibfnamefont{I.~A.} \bibnamefont{Campbell}},
  \bibinfo{author}{\bibfnamefont{K.}~\bibnamefont{Hukushima}},
  \bibnamefont{and} \bibinfo{author}{\bibfnamefont{H.}~\bibnamefont{Takayama}},
  \emph{\bibinfo{title}{Extended scaling scheme for critically divergent
  quantities in ferromagnets and spin glasses}}, \bibinfo{journal}{Phys. Rev.
  Lett.} \textbf{\bibinfo{volume}{97}}, \bibinfo{pages}{117202}
  (\bibinfo{year}{2006}).

\end{thebibliography}

\end{document}